\RequirePackage{fix-cm}
\documentclass[pdflatex,sn-basic]{sn-jnl}
\usepackage[most]{tcolorbox}
\usepackage{graphicx}
\usepackage{amsmath}
\usepackage{bookmark}
\DeclareFontShape{OMX}{cmex}{m}{n}{<-7.5>cmex7<7.5-8.5>cmex8<8.5-9.5>cmex9<9.5->cmex10}{}
\usepackage{manyfoot}
\DeclareNewFootnote{A}[gobble]
\usepackage{booktabs}
\usepackage{array}
\usepackage{colortbl}
\usepackage{tabularx}
\usepackage{url}
\usepackage{xcolor}
\usepackage{hyperref}

\begin{document}

\title[Merged, Not Measured]{Merged, Not Measured: An Empirical Study of Performance Issues Fixed by Coding Agents}

\author[1]{Zhenyu Qi}\email{qzydustin@arizona.edu}
\author[1]{Haotang Li}\email{haotangl@arizona.edu}
\author*[2]{Jinfu Chen}\email{jinfuchen@whu.edu.cn}
\author[3]{Huashan Chen}\email{chenhuashan@iie.ac.cn}
\author[4]{Yutong Zhao}\email{Yutong.Zhao@csulb.edu}
\author[5]{Derui Zhu}\email{dxzvse@rit.edu}
\author[1]{Tomas Cerny}\email{tcerny@arizona.edu}
\author[1]{Bo Liu}\email{boliu@arizona.edu}
\author*[1]{Sen He}\email{senhe@arizona.edu}
\affil[1]{\orgdiv{School of Electrical, Computing, and Software Engineering},
  \orgname{University of Arizona},
  \orgaddress{\city{Tucson}, \state{AZ}, \country{USA}}}
\affil[2]{\orgdiv{School of Computer Science},
  \orgname{Wuhan University},
  \orgaddress{\city{Wuhan}, \country{China}}}
\affil[3]{\orgdiv{Institute of Information Engineering},
  \orgname{Chinese Academy of Sciences},
  \orgaddress{\city{Beijing}, \country{China}}}
\affil[4]{\orgdiv{Department of Computer Engineering and Computer Science},
  \orgname{California State University, Long Beach},
  \orgaddress{\city{Long Beach}, \state{CA}, \country{USA}}}
\affil[5]{\orgdiv{Department of Software Engineering},
  \orgname{Rochester Institute of Technology},
  \orgaddress{\city{Rochester}, \state{NY}, \country{USA}}}
\date{Received: date / Accepted: date}

\abstract{Coding agents open pull requests (PRs) that claim to speed up software, but studies of human performance fixes say little about how maintainers respond to such a fix or whether its claim holds.
From the 71,677 agent PRs of AIDev v4, a text filter and codebook coding by language models and by the authors select 1,262 performance issues fixed by six agents in 582 repositories.
We code each issue and its tests and re-execute 23 rejected and 30 merged fixes.
(1) 57\% of closed fixes are merged, 61\% of rejections give no stated reason, and only 6 of the 23 re-executed rejected claims held under our three-run pilot on mostly agent-built workloads.
(2) Acceptance rises with the agent's track record in the repository (31--37\% to 70\%) and with the repository's pre-opening merge rate on its other agent PRs (33\% to 84\%).
Merged fixes delete a larger share of the lines they change (0.26 versus 0.15), a difference that holds within agent and within repository, with no such difference detected in the coded content, description, tests or measurements.
(3) Repeated computation and redundant data processing cause 44\% of the issues, and 46\% of fixes are architectural-level.
(4) Agents change tests in 37\% of fixes and 11\% carry a performance test or benchmark; of the 30 merged fixes, 18 met our delivery criterion, 3 fell short of the claim, 9 showed no significant gain or regressed, and 14 change behavior on untested inputs.
The outcome tracks the repository's history with the agent rather than the coded content of the fix, and a merge does not show that the fix delivers what it claims.
}
\keywords{coding agents, performance issues, code review, mining software repositories}

\maketitle

\section{Introduction}
\label{sec:intro}

Coding agents such as GitHub Copilot's coding agent, OpenAI Codex, Devin, Cursor, Claude Code and Google Jules now open pull requests (PRs) that fix performance issues in public repositories \citep{li2026aidev,msr2026codeopt,msr2026perfopt}, and the PR description often claims a speed-up.
A performance issue is a runtime inefficiency of the product (time, memory, I/O, queries, start-up) that a change to production code removes \citep{jin2012perfbugs,zhao2023perf}.
The AIDev dataset records 932,791 agent PRs at its MSR 2026 release \citep{li2026aidev} and 2.7 million at the v4 snapshot we use \citep{li2026aidevdata}, 71,677 of them in repositories over 100 stars.
However, a passing test cannot show the benefit of such a fix, so a maintainer who receives one must judge the benefit from a measurement, and a reliable measurement takes repetition and control that a PR thread rarely shows \citep{georges2007rigorous,mytkowicz2009wrongdata}.
A wrong decision then merges a regression or discards a real improvement.

Prior work has studied performance fixes and the acceptance of PRs separately.
Studies of human-written performance fixes describe them as small patches: a median of 8 changed lines \citep{jin2012perfbugs}, and 73\% of JavaScript optimizations under 20 lines \citep{selakovic2016js}.
\citet{nistor2013perfbugs} add that such fixes are nevertheless larger and harder to get right than other bug fixes, and \citet{zaman2012perfbugs} find that only about a third of performance reports carry a measurement.
\citet{zhao2023perf} coded 570 of these fixes into eight root causes and found 27\% architectural-level, 15\% with a test change, and maintainability or readability weighed against a fix in a minority of the issues.
Whether a PR is merged, in turn, has long been tied to who submits it.
\citet{tsay2014influence} showed that a submitter's prior interaction with the project and social ties to its maintainers raise the odds of acceptance more than tests do, and \citet{wyrich2021bots} found bot PRs merged about half as often as human ones.
For coding agents, the MSR 2026 studies of AIDev coded why agent PRs are rejected, pooling every kind of task \citep{msr2026rejections,msr2026boot,msr2026fixrejection,wang2026rejectedmore}, and \citet{duma2026reviews} found that most agent PRs receive no human review.
Two of these studies looked at performance PRs in particular: \citet{msr2026codeopt} selected them by AIDev's task label, without confirming each fix by hand, and found that agent PRs state a validation less often than human ones, and \citet{msr2026perfopt} selected them with a language-model classifier and found acceptance varying with the type of optimization in pooled comparisons.
Benchmarks such as SWE-Perf, GSO and SWE-fficiency measure the speed-ups agents achieve on real repositories under a fixed harness and find them far behind the experts' \citep{he2025sweperf,shetty2025gso,ma2025swefficiency}.
However, \textbf{no study of the performance fixes that agents open in real repositories confirms each fix by hand, separates the fix's content from the repository that received it, or re-executes what the fix claims.}
The studies of human fixes describe fixes that a project requested from an author it knows, the acceptance studies pool what a fix contains with who submitted it, and the benchmarks measure a speed-up in a harness rather than in the repository the agent submitted to.
This study asks how maintainers decide on a performance fix that an agent opens, and whether the reported improvement holds when the fix is run.

To answer it, we study the performance issues that agents fix in AIDev v4 as a population of their own.
Each performance issue is identified through the agent's PR that fixes it, and the agent's change is the issue's fix, so we count and code performance issues and use ``PR'' only for the GitHub object.
From the 71,677 agent PRs, a text filter and codebook coding, first by language models and then by the authors, select 1,262 performance issues fixed by six agents in 582 repositories (Section~\ref{sec:design}).
For each issue we code the inefficiency, the fix, its scope and its tests, and we re-execute 30 rejected fixes whose PRs report a measurement (23 ran) and 30 merged fixes that change a test, to test whether the claim holds on either side of the decision.
A comparison with human PRs in five of the repositories (Section~\ref{sec:funnel}) locates the acceptance gap between the two.
This study aims to answer the following research questions:

\smallskip\noindent\textbf{RQ1: How often are agents' performance fixes accepted or rejected?}
We find that 57\% of closed fixes are merged and 61\% of rejections state no reason, and that only 6 of 23 re-executed claims reproduced in our three-run pilot on mostly agent-built workloads.

\smallskip\noindent\textbf{RQ2: What factors are associated with the acceptance or rejection of agents' performance fixes?}
We find that the agent's track record in the repository and the repository's pre-opening merge rate on other agent PRs explain the outcome better than the other fix properties we measured; only the deleted-line share survives within agent and repository.

\smallskip\noindent\textbf{RQ3: How frequently do agents' performance fixes involve architectural changes?}
We find that agents fix the same inefficiencies as humans, most often by caching, but that 46\% of their fixes are architectural-level, with one or two functions changed per file, against 27\% of the human fixes \citet{zhao2023perf} coded in 13 projects.

\smallskip\noindent\textbf{RQ4: What evidence do agents provide to support their performance fixes?}
We find that agents change tests in 37\% of their fixes, that only 11\% carry a performance test or benchmark, and that acceptance is no higher with any kind of test.
Of 30 re-executed merged fixes, 18 met our delivery criterion, 3 improved below the claim, 9 showed no significant gain or regressed, and 14 change behavior on untested inputs.

\smallskip
This paper makes four contributions.

\begin{itemize}
\item Evidence that acceptance tracks the repository's history with the agent and the shape of the patch, while the coded content and evidence variables do not separate the outcomes.
\item A re-execution of 30 rejected fixes with a claimed improvement, of which 23 ran and 6 held, and of 30 merged fixes, of which 18 met our delivery criterion and 14 change behavior, with the protocol for both.
\item A characterization of the inefficiencies agents fix, the resolutions they apply and the reach of their changes.
\item A dataset of 1,262 agent performance fixes, coded for issue, fix, scope and tests, with the materials to reproduce our analyses.
\end{itemize}

Section~\ref{sec:background} reviews prior work, Section~\ref{sec:design} describes the dataset and the design of each research question, Sections~\ref{sec:rq1} to~\ref{sec:rq4} answer RQ1 to RQ4, and Sections~\ref{sec:discussion} to~\ref{sec:conclusion} draw implications, discuss threats and conclude.

\section{Background and Related Work}
\label{sec:background}

This section reviews the four strands of prior work on which the study rests: performance measurement, which sets the evidence a performance claim needs and supplies the protocol of our re-execution (Section~\ref{sec:bg:analysis}); how PRs from outsiders, bots and coding agents fare on GitHub, the acceptance literature that RQ2 separates from the content of a fix (Section~\ref{sec:bg:agents}); LLM-based optimizers and the benchmarks that score them by a measured speed-up, the score our study adds to the merge outcome (Section~\ref{sec:bg:optimizers}); and studies of human performance fixes, whose codes and text patterns are the instruments of RQ3, RQ4 and the dataset construction (Section~\ref{sec:bg:human}).

\begin{figure}[t]
\centering
\includegraphics[width=0.88\linewidth]{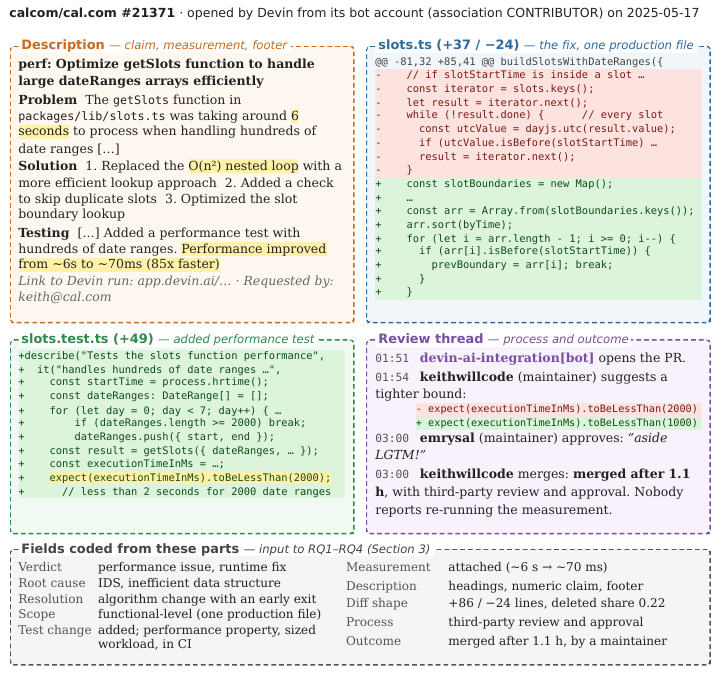}
\caption{Anatomy of an agent performance fix (calcom/cal.com PR 21371, Devin, merged): the PR description with its claim and attached measurement, the production change, the added performance test, the review thread with the decision, and the fields the study codes from each part (Section~\ref{sec:design}).}
\label{fig:anatomy}
\end{figure}

Figure~\ref{fig:anatomy} shows the object that the four strands meet in: one agent performance fix as it appears on GitHub, with the parts that the study reads.
In this example, a Devin bot account opened the fix, a maintainer suggested a tighter bound for the added performance test three minutes later, a second maintainer approved it, and the first merged it 1.1 hours after opening, without anyone reporting a re-run of the claimed speed-up.
Section~\ref{sec:design} describes how each of these parts is coded for the 1,262 fixes.

\subsection{Performance Analysis}
\label{sec:bg:analysis}

Measuring a performance change reliably takes repetition and control that a PR thread rarely shows.
Early on, \citet{georges2007rigorous} showed that reporting the best or the mean of a few runs of a Java benchmark can reverse a comparison, and they proposed confidence intervals over repeated runs in its place.
\citet{mytkowicz2009wrongdata} then showed that link order and the size of the process environment alone shift measured run time enough to reverse the sign of a claimed speed-up.
In the cloud, \citet{laaber2019cloud} measured coefficients of variation from 0.03\% to over 100\% across microbenchmarks and instance types.
A 10\% slowdown was detectable only when test and control ran on the same instance in randomized order.
More recently, \citet{traini2023steadystate} ran 586 JMH benchmarks from 30 Java systems and found that many never reach a steady state within the warm-up their developers configured.

Open-source projects apply little of this rigor.
In 111 Java projects that hold performance tests, a single developer wrote every test in 48\% of the projects and only 16\% used a benchmarking framework rather than repurposed unit tests \citep{leitner2017perftest}; performance unit testing lags functional testing \citep{stefan2017jmh}; and the suites of ten Java and Go projects took 11 minutes to 8.75 hours to run, with results varying by 50\% or more on bare metal \citep{laaber2018microbench}.
Regressions therefore enter unnoticed: \citet{chen2017perfregression} found that 19.6\% of Hadoop commits and 67.4\% of RxJava commits introduced at least one significant regression, and \citet{chen2020perfjit} later predicted such commits from code and history metrics at a fraction of the testing cost.
\citet{jangali2022ju2jmh} generated JMH benchmarks from unit tests and identified insufficient workload and unstable benchmarks as the two recurring defects of generated performance tests.
This literature sets what a performance claim needs.
Section~\ref{sec:pilot} borrows the smallest safeguards (repeated runs of one workload on both trees, one container) to re-execute rejected claims, and RQ4 codes whether an agent's added tests assert a performance quantity, and on what workload.

\subsection{LLMs and Agents for Software Engineering Tasks}
\label{sec:bg:agents}

Whether a contribution is merged on GitHub has been explained by its author's standing since before agents existed.
\citet{gousios2014pullbased} found that most PRs are decided within a day.
In the same year, \citet{tsay2014influence} showed on 659,501 PRs that the author's prior interaction with the project and social connection to its maintainers raise acceptance more than the presence of tests.
Later, \citet{li2022abandonment} studied abandoned PRs and found that the reason contributors give most often is silence from the integrators.
Bots then inherited the outsider's position.
\citet{wyrich2021bots} found bot PRs merged in 37\% of cases against 73\% for human PRs, and \citet{alfadel2021dependabot} found a third of Dependabot's security updates declined.
\citet{wessel2021disturb} and \citet{he2023dependabot} traced maintainers' perception of bot activity as noise.

Coding agents are evaluated first on benchmarks and then in the repositories they now contribute to.
\citet{jimenez2024swebench} defined SWE-bench, 2,294 GitHub issues whose fix is scored by hidden tests.
\citet{yang2024sweagent}, \citet{zhang2024autocoderover}, \citet{wang2025openhands} and \citet{xia2025agentless} then built the agent designs from which the commercial agents of our study descend.
Meanwhile, on GitHub, \citet{chouchen2024chatgpt} found that PRs written with ChatGPT close more slowly, and \citet{xiao2024prdesc} found 18,256 PRs with Copilot-generated descriptions merged more often and reviewed faster.
\citet{li2025teammates} and \citet{li2026aidev} then collected the AIDev dataset, the PRs that six coding agents opened on GitHub with their outcome, text and agent, and \citet{watanabe2025agentic} found that most Claude Code PRs are merged but nearly half only after human revision.
The MSR 2026 mining challenge produced studies of acceptance across all AIDev tasks.
\citet{msr2026rejections} inspected 717 closed agent PRs and found no observable rationale in 33\% of rejections, and \citet{msr2026boot} and \citet{msr2026fixrejection} coded the rejection reasons of agent PRs and of agent fixes.
\citet{wang2026rejectedmore} found AI-generated PRs rejected more often for incomplete implementation and inadequate testing, and \citet{duma2026reviews} found that most AI-generated PRs receive no human review.
Most recently, \citet{cynthia2026postmerge} found on 1,210 merged bug-fix PRs that a merge says little about the code quality that follows it.
These studies pool every kind of agent task, so the author-standing effect they inherit from the human studies has not been separated from what a particular kind of fix contains.
RQ2 makes that separation for one task by comparing merged and rejected fixes within agent and within repository, and Section~\ref{sec:funnel} locates the gap against human PRs in the same repositories.

\subsection{LLMs and Agents for Performance Optimizations}
\label{sec:bg:optimizers}

Benchmarks score an optimizer by the speed-up it delivers against a reference in a fixed harness.
\citet{garg2022deepdevperf} trained a transformer on developer-written C\# performance commits, \citet{shypula2024pie} built PIE from 77,000 pairs of competitive C++ submissions scored in a simulator, and \citet{huang2024effibench} found that GPT-4's solutions to 1,000 LeetCode problems run in 3.12 times the time of the reference solutions.
Later benchmarks added efficiency-aware pass metrics and file-level problems \citep{du2024mercury,qiu2025enamel,peng2025coffe}, and further studies found refinement raising efficiency moderately while lowering correctness \citep{waghjale2024ecco}, generated-code efficiency unrelated to a model's correctness or size \citep{niu2024efficiency}, and LLM solutions faster than most human LeetCode submissions \citep{coignion2024leetcode}.
\citet{li2026aicodeperf} traced the inefficiencies of Copilot, CodeLlama and DeepSeek-Coder output to four root causes, and \citet{gao2025sbllm} and \citet{peng2025perfcodegen} improved a generator's speed-ups with search and execution feedback.

Repository-level benchmarks bring the harness to real projects and find agents far behind experts.
\citet{he2025sweperf} scored 140 performance PRs from nine Python repositories.
Then \citet{shetty2025gso} found agents solving under 5\% of 102 optimization tasks, and \citet{ma2025swefficiency} measured agent speed-ups at under 0.23 of the expert's on 498 tasks.
In turn, \citet{garg2025perfbench} raised an agent from 3\% to 20\% on 81 .NET performance bugs by having it benchmark its own fix.
\citet{yi2025fastercode} ran LLM patches for 65 Java tasks against developer-written benchmarks and found about a third of the plausible patches without a measurable gain and the rest behind the human fix.
Two studies of AIDev select performance PRs by AIDev's task label \citep{msr2026codeopt} or by a language-model classifier \citep{msr2026perfopt}.
\citet{msr2026codeopt} found that 324 agent performance PRs state a validation less often than 83 human ones (45.7\% versus 63.6\%), and \citet{msr2026perfopt} grouped agent performance PRs into 52 topics and found acceptance varying by optimization type.
In a benchmark the score is a measured speed-up, and in a repository the score is the merge.
Our study takes the second score as its outcome and adds the first to it: it confirms each performance issue by codebook coding, first by language models and then by the authors (Section~\ref{sec:population}), asks in RQ2 whether the coded content of a fix or the repository it lands in separates the outcomes, and re-executes in RQ1 the claimed speed-ups of rejected fixes inside the repository the agent submitted them to.

\subsection{Human Fixes of Performance Issues}
\label{sec:bg:human}

Studies of human fixes describe performance issues as slow to resolve, small to patch, and verified by measurement in a minority of cases.
\citet{zaman2011secperf} found in Firefox that performance bugs are fixed more slowly than security bugs and touch 2.6 times as many files, and \citet{zaman2012perfbugs} then found that about a third of performance reports in Firefox and Chrome carry a measurement.
In the same period, \citet{jin2012perfbugs} studied 109 performance bugs in five C/C++ systems and found most patches small, while \citet{nistor2013perfbugs} found performance fixes larger than other fixes and the bugs discovered by reading code far more often than by profiling.
\citet{selakovic2016js} found that 98 JavaScript optimizations change fewer than 20 lines in 73\% of cases and that 16\% regress on at least one engine.
Other studies classified Android performance bugs, three quarters of them GUI lag \citep{liu2014android}, mined performance commits from Android apps \citep{das2016android}, had detected memoization opportunities confirmed by projects' developers \citep{dellatoffola2015memoization}, tied 59\% of 193 performance bugs to a configuration option \citep{han2016configperf}, found profiling data in 41\% of 192 issues \citep{zhao2020icpe}, and measured significant execution-time change in 55\% of refactoring commits \citep{traini2022refactoring}.
\citet{zhao2023perf} coded 570 issues from 13 projects into eight root causes (Table~\ref{tab:rootcause}), classified each fix as localized (one file) or design-level (a coordinated revision of a group of related files), which we call functional-level and architectural-level, and found 27\% of fixes architectural-level and 15\% accompanied by test changes.
\citet{zhao2024hlp} then derived heuristic linguistic patterns (HLPs) that identify performance issue reports from their text.
These two studies supply our instruments: the root-cause codes and the functional-level/architectural-level distinction are the vocabulary of RQ3, and the HLPs are the text filter of Section~\ref{sec:population}.
RQ3 and RQ4 measure for agent fixes the properties these studies measured for human fixes: root cause, resolution, scope and test content.
Their corpora were drawn from issue reports in projects they chose and ours from agent PRs, so the shares describe two populations rather than one population under two authors.

\section{Study Design}
\label{sec:design}

\begin{figure}[t]
\centering
\includegraphics[width=0.88\linewidth]{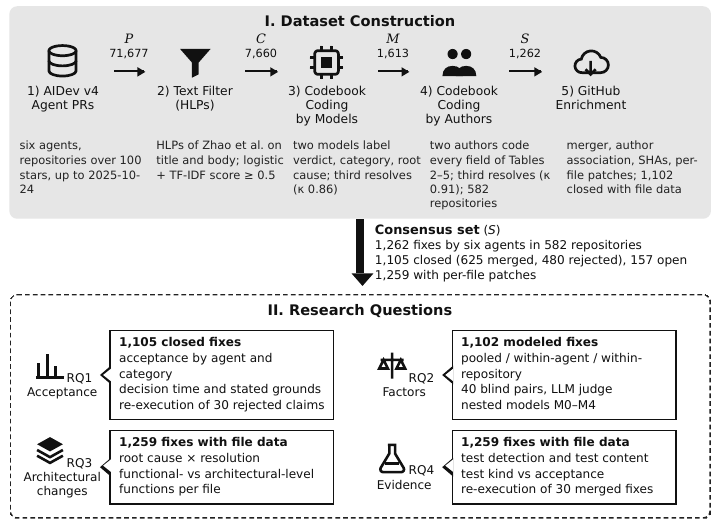}
\caption{Overview of the study. Top: the five steps that reduce the 71,677 agent PRs of AIDev v4 to the consensus set of 1,262 performance issues (\(P\), \(C\), \(M\) and \(S\) are the sets of Table~\ref{tab:flow}). Bottom: the four research questions with the subset of fixes and the analyses each one uses.}
\label{fig:workflow}
\end{figure}

Figure~\ref{fig:workflow} shows the overview of our study.
We start from the agent PRs of AIDev v4, keep the PRs whose text describes a performance fix, and confirm each of them by coding against a written codebook (\emph{codebook coding}), first by language models and then by the authors, who also code the issue, its fix and its tests.
On the resulting dataset we answer the four research questions of Section~\ref{sec:intro}.
Section~\ref{sec:population} describes the construction of the dataset, and Sections~\ref{sec:design:rq1} to~\ref{sec:design:rq4} give the design of each research question with the analysis we run and the metrics we report.

\subsection{Dataset Construction}
\label{sec:population}
\label{sec:labeling}
\label{sec:coding}

To construct the dataset, we reduce the 71,677 agent PRs of AIDev v4 \citep{li2026aidevdata} in three steps, a text filter, codebook coding by language models and codebook coding by the authors, which leave 7,660, 1,613 and 1,262 PRs in turn (Table~\ref{tab:flow}).

\begin{table}[t]
\caption{From AIDev v4 to the analysis sets.}
\label{tab:flow}
\small
\setlength{\tabcolsep}{4pt}
\begin{tabularx}{\textwidth}{>{\raggedright\arraybackslash}Xr>{\raggedright\arraybackslash}X}
\toprule
Set & n & Used in \\
\midrule
\rowcolor{gray!12} Agent PRs in repositories over 100 stars & 71,677 & repository merge rate (Sec.~\ref{sec:rq2:context}) \\
Passed the text filter (score $\ge$ 0.5) & 7,660 & input to the model coding \\
\rowcolor{gray!12} Model-positive set (codebook coding by models) & 1,613 & input to the author coding; sensitivity check \\
Consensus set (codebook coding by the authors) & 1,262 & all main tables; 582 repositories \\
\rowcolor{gray!12} \quad closed (merged or rejected) & 1,105 & acceptance (Secs.~\ref{sec:rq1}, \ref{sec:rq2}) \\
\quad with per-file patches & 1,259 & scope, tests (Secs.~\ref{sec:rq3}, \ref{sec:rq4}) \\
\rowcolor{gray!12} \quad rejected, with a measurement, re-executed & 30 $\rightarrow$ 23 & re-executed claims (Sec.~\ref{sec:rq1:validity}) \\
\quad merged, with a test change, re-executed & 30 & delivered improvement (Sec.~\ref{sec:rq4:delivered}) \\
\bottomrule
\end{tabularx}
\end{table}

\smallskip\noindent\textbf{Source data.}
AIDev v4 holds 71,677 PRs opened by six agents (Copilot, Codex, Devin, Cursor, Claude Code and Jules) in repositories with more than 100 stars, up to 2025-10-24.
Each PR carries AIDev's state at that snapshot: merged, open, or closed without a merge, which we call \emph{rejected}.
A rejection is a recorded state, and Section~\ref{sec:rq1:process} separates closures an inactivity mechanism produced from those a person made.

\smallskip\noindent\textbf{Step 1.}
\textit{To find candidate performance fixes among the 71,677 PRs, we apply the method of \citet{zhao2024hlp} to the words of each PR's title and body.}
\citet{zhao2024hlp} identify performance issue reports from their text with 80 heuristic linguistic patterns (HLPs), which they derived from the issue reports of 13 Apache projects over 13 rounds of manual tagging until no new pattern appeared.
An HLP is a rule that matches one sentence: lexical rules (44\% of the 80) match performance words such as ``takes too long'' or ``high CPU'', structural rules (38\%) match a sentence's shape such as ``byte by byte'', and semantic (10\%) and profiling (8\%) rules match a sentence's meaning or pasted profiler output.
Together, on their data the HLPs find performance sentences with 94\% precision and 87\% recall.
We reproduce the 80 HLPs as rules and run them over the sentences of each PR's title and body, after removing code blocks, links, PR templates and pasted shell commands.
The HLPs were written for issue reports, which describe a symptom, whereas an agent PR describes its fix, so we add 28 patterns for fix vocabulary (memoize, debounce, lazy load, batch, N+1 query) and three for categories the codebook excludes (CI speed-ups, generated code, benchmark infrastructure).
Two logistic regressions then turn the text into a score, one over the matched patterns (a title indicator and a log count per pattern) and one over TF-IDF features of the title and cleaned body; the screen score is the mean of their probabilities.
Both were trained on 1,219 PRs labeled in an earlier round of this project plus 8,000 weak negatives (unlabeled PRs without a performance keyword in the title), and the labeled PRs serve only as training data for the screen.
In five-fold cross-validation on AIDev v3, a score of at least 0.5 recovers 80\% of the labeled fixes while flagging 1.7\% of unlabeled PRs, and a random audit of PRs under the cut (50 with a score in [0.3, 0.5) and 100 below 0.3 in v3, and 50 in [0.3, 0.5) among the later v4 PRs) found one, zero and two performance fixes, which puts the screen's recall against the population at roughly 90\%.
We keep the 7,660 PRs (10.7\%) whose score is at least 0.5 as candidates; the 94\% precision and 87\% recall above are Zhao et al.'s figures for the original HLPs on issue reports, not a measurement of this screen.

\smallskip\noindent\textbf{Step 2.}
\textit{Language models label each of the 7,660 candidates by codebook coding.}
However, reading every candidate by hand is beyond the reach of three authors, so at this step language models carry out the coding procedure that the authors apply at Step 3, and each model sees the PR's title, body, task type, changed lines and most-changed files.
The codebook defines a \emph{performance issue} operationally as a runtime inefficiency of the product (time, memory, I/O, queries, start-up) that the PR's main change to production code aims to remove.
It keeps a candidate when the PR fixes such an issue, and it excludes CI speed-ups, compiler output, benchmark-only PRs and features whose bodies mention performance.
Two models (Claude Sonnet and GPT-5.6) label each candidate independently with one of three verdicts (performance issue, performance-related other, or not performance), the issue's performance category and its root cause, the first three fields of Table~\ref{tab:codes}.
A verdict on which the two agree is accepted, and a third model (Gemini 3.5 Flash) resolves each disagreement.
The identifiers, versions, decoding settings and prompts of every model instrument in this study (the coding models here, the repository-category models of Section~\ref{sec:design:rq1:approach}, and the judge and executing agents of RQ1, RQ2 and RQ4) are listed in the replication package.
The two models agree on the verdict with a Cohen's \(\kappa\) of 0.86 \citep{cohen1960kappa}, which \citet{landis1977kappa} class as almost perfect agreement.
For two coders, Cohen's \(\kappa\) compares the observed share of items on which they agree, \(p_o\), with the share expected from their marginal code frequencies, \(p_e\), as \((p_o - p_e)/(1 - p_e)\), and we read each \(\kappa\) on the scale of \citet{landis1977kappa}.
The paper reports \(\kappa\) for the verdict of each coding stage and for the repository category, and the replication package holds both coders' codes for every field.
As a result, the 1,613 candidates whose final verdict is a performance issue form the \emph{model-positive set}.
For the 6,047 candidates the models rejected, three authors audited every tenth candidate (604) by the Step 3 coding procedure; the two coders agreed with a \(\kappa\) of 0.96 and confirmed the rejections.

\smallskip\noindent\textbf{Step 3.}
\textit{Three of the authors then code every fix in the model-positive set against the coding schemes of Tables~\ref{tab:codes} and \ref{tab:codes:rq1} to~\ref{tab:codes:rq4}.}
Coding a fix from its PR text, diff and review thread involves judgment, so we follow the two-step coding procedure of prior work \citep{li2026aicodeperf} to limit subjectivity.
Each field of these tables is coded in a pass of its own, with the codes of \citet{zhao2023perf} as the starting scheme where they apply; each pass has two steps.
\emph{Dual coding.} Two authors independently code the field for every fix, and a code on which they agree is accepted.
\emph{Disagreement resolution.} The third author reads the fix and both codes and decides disagreements against the codebook.

\begin{table}[!t]
\caption{Coding scheme of the dataset. Each row is one coding pass: two coders coded the field independently for every fix it applies to, and a third resolved their disagreements.}
\label{tab:codes}
\small
\setlength{\tabcolsep}{4pt}
\begin{tabularx}{\textwidth}{>{\raggedright\arraybackslash}p{2.4cm}>{\raggedright\arraybackslash}p{1.9cm}>{\raggedright\arraybackslash}X}
\toprule
Field & Applies to & Codes \\
\midrule
\rowcolor{gray!12} Verdict & every fix & performance issue; performance-related other; not performance \\
Category & every fix & runtime; memory; concurrency; I/O, network or database; start-up or load \\
\rowcolor{gray!12} Root cause \citep{zhao2023perf} & every fix & repeated computation (RC); redundant data processing (RDP); general inefficient computation (GIC); inefficiency under special cases (ISC); inefficient API usage (IAU); inefficient synchronization (IS); inefficient data structure (IDS); inefficient iteration (II); unclear \\
Attached measurement & every fix & yes (the PR text reports a timing, memory figure, benchmark or profiler output); no \\
\bottomrule
\end{tabularx}
\end{table}
The two coders agree on the verdict with a \(\kappa\) of 0.91, and the resolved verdicts confirm 1,262 of the 1,613 fixes; these 1,262 performance issues in 582 repositories form the \emph{consensus set} on which every main table reports.
Of the 351 fixes the coders did not confirm, 270 are features or refactors that mention performance.
The model-positive set is a sensitivity check: its acceptance, 58\%, is close to the consensus 57\%.

\smallskip\noindent\textbf{Enrichment.}
By agent, the consensus set holds 488 Copilot, 438 Codex, 147 Devin, 112 Cursor, 60 Claude Code and 17 Jules fixes, of which 625 are merged, 480 rejected and 157 open.
We split repository size into star terciles (small \(\le\) 347, mid 348--2,576, large \(\ge\) 2,577).
The agents differ in which account opens their PRs, and hence who merges them; we call this the \emph{account model}.
For Copilot, Devin and Jules, the PR account is a bot account, so someone else makes each merge, whereas for Codex, Cursor and Claude Code, the PR account is the human operator, who usually merges the fix.
RQ1 and RQ2 therefore report the two groups separately.
AIDev records the outcome, text and agent of each PR but not the merger, the author association, the commit SHAs or the per-file patches, so we fetched these fields from GitHub for 1,527 PRs and used AIDev's commit details for the 86 that returned 404.
Per-file patches are missing for three fixes, so analyses over file data cover 1,259 fixes (1,102 closed).

\subsection{RQ1: Acceptance and Rejection of the Fixes}
\label{sec:design:rq1}

\subsubsection{Analysis Design}
\label{sec:design:rq1:approach}
\label{sec:pilot}

To answer RQ1, we describe how the 1,105 closed fixes of the consensus set were decided, in three steps.
First, we compute the overall acceptance rate and break it down by the category of the receiving repository, by agent, and by the two together, with star tercile, the account model, and the author association that GitHub records for the PR account (OWNER, MEMBER, CONTRIBUTOR or NONE) as further splits.
GitHub records NONE when the PR author has no recognized association with the repository, and CONTRIBUTOR, MEMBER and OWNER mark progressively stronger relationships.
In our data a bot account is recorded as NONE until the repository has merged one of its PRs, after which it appears as CONTRIBUTOR, so for agent accounts NONE coincides with an empty track record.
Because two repositories contribute 80 and 41 fixes to one category each, we recompute the category rates without them as a sensitivity check.
We also report the rate by month of opening, count each repository once to give the share of repositories in a category that merged every fix or none, and leave out cells with fewer than ten fixes.
The repository category follows the ``Apache category'' of \citet{zhao2023perf}, which takes each project's domain from the Apache Software Foundation project directory: their eight labels (Content, Library, Big-data, Build Management, Server-client, Cloud, Search and Language) first, the directory's other labels (for example database, web-framework, mobile or security) when none of the eight fits, and Other for projects with no ASF analogue.
Following their convention, a library with a specific domain takes the domain as its primary category and Library as a secondary one.
Two independent model passes, on different models, labeled each of the 582 repositories from its GitHub description, topics and homepage under a written codebook (Table~\ref{tab:codes:rq1}).
They agreed on 85\% of the primary labels (\(\kappa\) = 0.83), and a third pass resolved the 88 disagreements against the codebook.
Second, we measure the time from opening to decision and separate rejections by how they close.
Third, we read the 480 rejections' stated grounds from the coded closing reason (Table~\ref{tab:codes:rq1}), and we re-execute a sample of rejected fixes whose PR text reports a measurement, to test whether the claim holds outside the PR thread.

For the re-execution, we drew 30 rejected fixes that report a measurement in their PR text, in TypeScript, JavaScript, Python, Go or Rust, with base and head commits in the snapshot (seed 7, stratified by language).
Seven of them could not be run, because three need a live database and four are whole-workspace Rust builds, so the shares below are over 23 fixes.
For each fix we check out the base commit and the PR's head commit, which GitHub keeps under a pull-request reference after the branch is deleted, and build both in a language container (2 CPUs, 4--6 GB, 30 minutes); a Claude Opus agent with shell access carries out the run from the repository, the two commit ids, the PR text and the claimed numbers, without the PR's outcome.
We run one workload on both trees: the workload the PR names where it names one, otherwise a benchmark or test that the fix adds, otherwise a small workload constructed on the changed path (16 of the 23 fixes), three repetitions per tree, and we record the median ratio of base to head cost.
For example, for a fix that claimed a 2--7\(\times\) speed-up of a path-deduplication function, the agent timed the function on both trees at 10, 200 and 1,000 input paths.
We also run the tests the fix touched on the head and record whether they pass.
The verdict follows Eq.~\ref{eq:repro} below.
The merged fixes that RQ4 re-executes share the build step and the container with this pilot but are measured at three workload sizes with 12 forks each and judged by a stricter rule (Section~\ref{sec:design:rq4:approach}), so the two arms are described in parallel and are not compared as one experiment.

\begin{table}[!t]
\caption{Coding scheme for RQ1, coded as in Section~\ref{sec:population}; the repository category was coded by language models (see text).}
\label{tab:codes:rq1}
\small
\setlength{\tabcolsep}{4pt}
\begin{tabularx}{\textwidth}{>{\raggedright\arraybackslash}p{2.4cm}>{\raggedright\arraybackslash}p{1.9cm}>{\raggedright\arraybackslash}X}
\toprule
Field & Applies to & Codes \\
\midrule
\rowcolor{gray!12} Closing reason & rejected fixes & superseded; regression or incorrect; performance benefit doubted; scope or noise; design rejected; duplicate; inactivity; withdrawn by author; CI or conflict; unclear \\
Measurement requested; human engaged & rejected fixes & yes or no; yes or no \\
\rowcolor{gray!12} Repository category \citep{zhao2023perf} & every repository & Content; Library; Big-data; Build Management; Server-client; Cloud; Search; Language; Other ASF category; Other \\
\bottomrule
\end{tabularx}
\end{table}
\subsubsection{Evaluation Metrics}
\label{sec:design:rq1:metrics}

Acceptance is the merged share of the closed fixes,
\begin{equation}
\mathit{Acc} = \frac{N_{\mathrm{merged}}}{N_{\mathrm{merged}} + N_{\mathrm{rejected}}},
\label{eq:acc}
\end{equation}
where open fixes enter neither the numerator nor the denominator.
RQ2 and RQ4 use the same outcome.
Time to decision is the number of hours from the PR's creation to its close, and we report medians with interquartile ranges (IQR).
We call a fix \emph{silent} when it closes with no human text after any duration, a silent rejection within 24 hours a \emph{direct rejection}, and a rejection after seven days or more a \emph{prolonged rejection}.
For the pilot, let \(r_{\mathrm{c}}\) be the improvement the PR claims and \(r_{\mathrm{m}}\) the improvement we measure.
Both are ratios of the base tree's cost to the head tree's cost on the same workload, and a claimed speed-up of \(2\times\) gives \(r_{\mathrm{c}} = 2\).
A claim counts as \emph{reproduced} when the measured improvement reaches half of the claimed one,
\begin{equation}
r_{\mathrm{m}} - 1 \;\ge\; \tfrac{1}{2}\,(r_{\mathrm{c}} - 1),
\label{eq:repro}
\end{equation}
and a claim without a number counts as reproduced when \(r_{\mathrm{m}} \ge 1.2\).
A claim is \emph{partly reproduced} when the head is faster than the base but below this bound, and \emph{not reproduced} when the head shows no improvement or a regression.
Both thresholds are operational choices of this study.
We compare shares with Fisher's exact test, multi-category distributions with \(\chi^2\) and Cramér's V, and durations with the Mann--Whitney U test; the three serve the remaining research questions too.

\subsection{RQ2: Factors Associated with Acceptance}
\label{sec:design:rq2}

\subsubsection{Analysis Design}
\label{sec:design:rq2:approach}

To answer RQ2, we contrast merged and rejected fixes in four comparisons that hold progressively more fixed, and we end with nested regression models.
The pooled comparison contrasts the 1,105 closed fixes on the coded content (root cause, category, resolution and scope), on the tests and the attached measurement, on the size and locality of the diff, and on the process (a review or comment from an account other than the PR's, a \emph{third-party review}, and an approving review from such an account, a \emph{third-party approval}).
Next, the within-agent comparison repeats the contrast inside Copilot, Codex and Devin.
Each agent writes descriptions in its own template, so the pooled data mix templates with outcomes, and only a within-agent contrast separates the two.
It covers the description and diff-shape features that blocks F1 and F2 of Section~\ref{sec:design:rq2:metrics} list.
Then, the within-repository comparison contrasts the merged and rejected fixes of the 74 repositories that hold both outcomes, so that the maintainers, the project conventions and the repository's history with agent PRs are the same on both sides.
The blind-pair comparison draws 40 pairs of one merged and one rejected fix from the same repository and agent, one pair per repository (Copilot 16, Codex 14, Cursor 4, Devin 3, Claude Code 2 and Jules 1).
A Claude Opus judge reads the title, description, file list and diff excerpts of both fixes in random A/B order with every outcome field removed.
For each pair the judge records the change style of each fix (replaces, adds, mixed), which fix adds more machinery, the reading cost of each (1--5) and which is more readable, the description quality of each (accurate, overclaims, vague, mismatch), which fix it would merge, and which it predicts was merged.
Within the same pairs we also check whether the deleted-line share, the diff size and the three coded flags of added machinery point to the merged fix, so that the judge's reading can be compared with the measured features.
Finally, the last group of variables describes the repository's context.
It comprises the author association, the repository's merge rate on its other agent PRs, the number of agent PRs in the repository, and the agent's track record in the repository before the fix opened.
To test where the fix itself matters, we split the fixes into three bands of the pre-opening merge rate (at or under 0.40, 0.40 to 0.91, and over 0.91) and compare the deleted-line share by outcome within each band.
Finally, we repeat the within-agent contrast on the silent decisions alone, the fixes that Copilot, Codex and Devin had merged or rejected with no human text, because a silent decision cannot rest on anything a reviewer wrote.
\subsubsection{Evaluation Metrics}
\label{sec:design:rq2:metrics}

The deleted-line share of a fix is the fraction of its changed lines that it deletes,
\begin{equation}
d = \frac{L_{\mathrm{del}}}{L_{\mathrm{add}} + L_{\mathrm{del}}},
\label{eq:delshare}
\end{equation}
so a fix that rewrites code in place has a share near 0.5 and a fix that only adds code has a share near 0 (Section~\ref{sec:rq2:shape} gives one fix of each kind).
The repository merge rate of a fix \(p\) in repository \(R\) is the merged share of the agent PRs decided before \(p\) opened,
\begin{equation}
\mathit{MR}(p) = \frac{\bigl|\{\,q \in D_R(p) : q \text{ is merged}\,\}\bigr|}{\bigl|D_R(p)\bigr|},
\label{eq:mr}
\end{equation}
where \(D_R(p) = \{q \in A_R \setminus \{p\} : t_{\mathrm{closed}}(q) < t_{\mathrm{open}}(p)\}\) and \(A_R\) is the set of agent PRs in \(R\) among the 71,677 PRs, so the rate uses only decisions a submitter could have seen.
The rate is undefined when no other agent PR was decided before \(p\) opened, and the models carry an indicator for that absence.
A snapshot variant computed over all of \(A_R \setminus \{p\}\), including open PRs and decisions made after \(p\) opened, serves as a robustness check (Section~\ref{sec:rq2:context}).
The track record of \(p\) counts the PRs of the same agent that \(R\) merged before \(p\) opened,
\begin{equation}
\mathit{TR}(p) = \bigl|\{\,q \in A_R : \mathrm{agent}(q) = \mathrm{agent}(p),\ q \text{ merged before } t_{\mathrm{open}}(p)\,\}\bigr|,
\label{eq:tr}
\end{equation}
so it uses only information available when the fix was submitted.
For the within-repository comparison we use a Wilcoxon signed-rank test over the per-repository differences, and for the blind pairs a binomial test against the chance rate of one half.
For the acceptance models, we fit logistic regressions on the 1,102 closed fixes with file data, cluster standard errors by repository, and report odds ratios (OR) with 95\% confidence intervals and McFadden's pseudo \(R^2\), \(1 - \ln\mathcal{L}_{\mathrm{model}} / \ln\mathcal{L}_{\mathrm{null}}\) against the intercept-only model \citep{mcfadden1974logit}.
The models test the four groups of factors that the comparisons of Section~\ref{sec:design:rq2:approach} single out, and each group enters as one block so that its contribution to the fit can be read from the change in pseudo \(R^2\).
Model M0 holds the control variables, which are the root cause, the attached measurement, locality, a test change, diff size (log lines), third-party review and repository stars (log).
Next, block F1 is the agent and its description style.
It holds the agent as a factor, so that every later coefficient is a within-agent contrast, together with the description features (length, template sections, hype rate, numeric claims, an issue reference and the agent's footer).
Block F2 is the shape of the diff, with the deleted-line share (Eq.~\ref{eq:delshare}) as its main variable and the file count, the noise files and the share of churn in the largest file beside it.
Block F3 is added machinery.
It holds whether the fix adds a production file, whether the coders judged it architectural-level, and whether its pattern is a new shared abstraction, which are the three coded forms of a fix that installs a mechanism.
Finally, block F4 is the repository's history with agent PRs.
It holds the author association, the pre-opening merge rate (Eq.~\ref{eq:mr}) and the number of agent PRs in the repository; the same-agent track record (Eq.~\ref{eq:tr}) measures the same history and is collinear with the rate, so it enters as a variant of the full model rather than beside the rate.
M1 to M4 add the four blocks to M0 in this order, so that the contribution of each block is read against the blocks before it.
Each block is tested with a likelihood-ratio test against the previous model, and we also drop each block from the full model in turn.
To separate the agent's identity from its prose, we further fit M0 with the agent alone.
The band comparison of Section~\ref{sec:design:rq2:approach} enters a logistic model of acceptance on the deleted-line share, the diff size and the band.
The models condition on closure by the snapshot and so exclude the 157 open fixes.

\subsection{RQ3: Content and Reach of the Fixes}
\label{sec:design:rq3}

\subsubsection{Analysis Design}
\label{sec:design:rq3:approach}

To answer RQ3, we describe each fix in four ways, cross the root cause with the resolution to find the dominant resolution of each cause, and relate each description to acceptance.
First, the root cause is the inefficiency the agent found, coded under the eight categories of \citet{zhao2023perf} (Table~\ref{tab:rootcause}), from repeated computation to inefficient iteration, plus an unclear code.
Second, the resolution is the mechanism the fix applies, coded under the fifteen codes of Table~\ref{tab:codes:rq3}.
For example, a fix that stores a computed value and reuses it is coded as caching or memoization, and a fix that turns a per-row query in a loop into one query is coded as batching.
Third, the scope separates a \emph{functional-level} fix, which changes at most one production file, from a multi-file fix.
However, a file count over-states architectural-level work, because a lock file or a regenerated bundle makes a one-file fix look like a multi-file one.
The coders read every fix with at least two production files (943 in the model-positive set, 721 in the consensus set) and judged whether its production files change together for one performance mechanism.
Such a fix is \emph{architectural-level}, and for each of them the coders recorded one of the five patterns of Table~\ref{tab:codes:rq3}.
Before the coded scope, we also report the raw count of production files (one, two, or three or more) with the lines, functions, tests and measurements of each group, so that the coded share can be read against the file count.
Fourth, the depth of a fix is how many functions it touches in how many files, computed from the per-file patches and compared by outcome.

\begin{table}[!t]
\caption{Coding scheme for RQ3, coded as in Section~\ref{sec:population}.}
\label{tab:codes:rq3}
\small
\setlength{\tabcolsep}{4pt}
\begin{tabularx}{\textwidth}{>{\raggedright\arraybackslash}p{2.4cm}>{\raggedright\arraybackslash}p{1.9cm}>{\raggedright\arraybackslash}X}
\toprule
Field & Applies to & Codes \\
\midrule
\rowcolor{gray!12} Resolution (up to three, first is primary) & every fix & cache or memoize; data-structure change; algorithm change; early exit or fast path; loop restructure; API replacement; batching or coalescing; remove redundant work; lazy or deferred; concurrency change; I/O or query tuning; resource release or bounding; allocation reduction; parameter tuning; other \\
Scope \citep{zhao2023perf} & $\ge$ 2 production files & architectural-level (the files change together for one mechanism); functional-level plus noise (lock files, generated output, formatting, unrelated edits); mixed purpose (bundled with unrelated work) \\
\rowcolor{gray!12} Pattern \citep{zhao2023perf} & architectural-level fixes & change propagation (a changed interface and its callers); optimization clone (the same change at several sites); parallel optimization (independent changes in one fix); classic design pattern; new shared abstraction (a cache, pool or helper used from several places; added in this study) \\
\bottomrule
\end{tabularx}
\end{table}
\subsubsection{Evaluation Metrics}
\label{sec:design:rq3:metrics}

We report each code as a share of the consensus set, pooled and by agent, and test the distribution across agents and across outcomes with \(\chi^2\) and Cramér's V.
The architectural-level share is the fraction of fixes with file data that the coders judged architectural-level, \(N_{\mathrm{architectural\text{-}level}} / N_{\mathrm{file\ data}}\), so a multi-file fix whose extra files are noise counts toward the denominator only.
A fix that changes one source file plus a lock file and a generated bundle is functional-level under this count.
We count the functions a fix touches as the distinct hunk-header contexts in its production patches, so the count is a lower bound.
For each architectural-level fix we report functions per file (functions touched over production files changed), together with production lines changed, hunks, lines per function, the deleted-line share (Eq.~\ref{eq:delshare}) and the share of fixes that add a production file, as medians by pattern.
Differences between patterns use the Kruskal--Wallis test, and differences by outcome the Mann--Whitney U test.

\subsection{RQ4: Tests in the Fixes}
\label{sec:design:rq4}

\subsubsection{Analysis Design}
\label{sec:design:rq4:approach}

To answer RQ4, we read the evidence a fix carries from the tests it changes, in three parts: how often and in what form agents change tests, what the changed tests check and whether that relates to acceptance, and whether a test predicts that the claimed improvement holds when the fix is re-executed.
We first detect whether a fix changes at least one test or benchmark file from the path of each changed file, over the 1,259 fixes with file data, and count the share pooled and by agent.
For the touched tests we report two granularities: the status of each touched test file in the per-file patches (added, modified, removed or renamed), and the coders' reading of each fix, which counts test cases and records whether the fix adds cases, modifies existing ones, does both, or removes them (Table~\ref{tab:codes:rq4}).
Then, for each fix with a test change, the coders also recorded what the changed tests exercise, the workload they run and whether they run in CI.
For example, a mock that asserts one call instead of \(N\) exercises a performance property, whereas a renamed symbol in six golden files is an adaptation.
We call a test of the performance property itself, or a benchmark that exercises it without an assertion, a \emph{performance test or benchmark}, and we report the two separately where the distinction matters; we group fixes into four kinds (a performance test or benchmark, functional tests only, adaptation only, and no test change).
We then compare acceptance across these kinds, pooled, within each agent, and with and without third-party review, and we refit model M0 with the four kinds in place of the test indicator.
Finally, we re-execute 30 merged fixes that change a test, so that the tests a merged fix carries can be read against the improvement it delivers.
We drew them from the 216 merged consensus fixes that change at least one test file.
Each fix has to be rebuilt and measured at its pre-merge and merged commits, so the sample is purposive: a fix was eligible when a deterministic in-process harness can drive its performance-relevant change (no network, GPU, user interface or database) and its project builds with the toolchains we had, which excluded Android apps, very large builds and C\# projects.
We ranked 30 main cases and 10 reserves across five languages; one main case (SkBlaz/py3plex PR 100) adds a new vectorized API beside the old code rather than replacing a code path, so no pre-merge version computes the same function, and a reserve replaced it.
The 30 fixes are Java 11, Python 7, TypeScript 6, Go 4 and Rust 2.
For each fix we build the base and the head and run the same operation at three workload sizes: the size of the PR's own test (S), a workload one to two orders of magnitude larger (M), and the size of the PR's claim or the largest that keeps one run of the base under a few seconds (L).
Each size runs in 12 forks per version, interleaved base--head--head--base in a controlled experiment environment that avoids resource contention, and we record the median cost per operation, the ratio of base to head with a bootstrap 95\% confidence interval, a Mann--Whitney U test and Cliff's \(\delta\).
We also run the PR's own tests on both trees and record whether they pass on the base, so that a test that passes on the unfixed code is known not to guard the improvement.
To see whether the head behaves as the base did, we generate tests for every changed line by slicing the diff, repair them until they compile, run them on both trees, and record every input on which the outputs differ.

\begin{table}[!t]
\caption{Coding scheme for RQ4, coded as in Section~\ref{sec:population}.}
\label{tab:codes:rq4}
\small
\setlength{\tabcolsep}{4pt}
\begin{tabularx}{\textwidth}{>{\raggedright\arraybackslash}p{2.4cm}>{\raggedright\arraybackslash}p{1.9cm}>{\raggedright\arraybackslash}X}
\toprule
Field & Applies to & Codes \\
\midrule
\rowcolor{gray!12} Test change & changes a test file & added; modified; both; removed \\
Tests exercise (any of) & changes a test file & functional behavior; performance property (time, memory, allocations, calls or queries, renders); adaptation only (renamed symbol, updated snapshot); regression guard (reproduces the inefficiency); benchmark without assertion \\
\rowcolor{gray!12} Workload; runs in CI & changes a test file & none, toy, sized or parameterized; yes or no \\
\bottomrule
\end{tabularx}
\end{table}
\subsubsection{Evaluation Metrics}
\label{sec:design:rq4:metrics}

The test co-change rate is the share of fixes with file data that change a test or benchmark file, \(N_{\mathrm{test\ change}} / N_{\mathrm{file\ data}}\), and the performance-test share replaces the numerator with the number of fixes that carry a performance test or benchmark, where a fix counts once even when it adds several.
Acceptance by test kind uses Eq.~\ref{eq:acc} within each kind, differences between kinds use Fisher's exact test and \(\chi^2\), and the refitted model reports an odds ratio per kind against fixes with no test change.
For a re-executed merged fix, the verdict is read at the primary workload, the largest of the three sizes (L).
The fix \emph{delivers} when the head beats the base by at least 5\% in median cost with Mann--Whitney \(p < 0.05\) and \(|\delta| \ge 0.147\), and, when the PR claims a number, the confidence interval of the ratio reaches it; it is \emph{below the claim} when the head is faster by these criteria but the interval stays under the claimed number; and it \emph{does not deliver} when the change is below 5\%, not significant, or in the wrong direction.
The merged arm is not compared statistically with the rejected arm, because the two differ in sampling, repetitions and verdict rule; within the merged arm we compare the fixes with and without a performance assertion with Fisher's exact test.
With 9 and 21 fixes in the two groups, the comparison detects only a large difference, so we report the counts.

\section{RQ1: How often are agents' performance fixes accepted or rejected?}
\label{sec:rq1}

\begin{tcolorbox}[enhanced, colback=gray!10, colframe=black!75, boxrule=0.8pt, arc=3pt, drop shadow={black!50!white}, left=5pt, right=5pt, top=3pt, bottom=3pt]
\textbf{Insight:} Maintainers decide on agents' performance fixes quickly and mostly without stating a reason, and most claimed improvements failed when we re-executed rejected fixes.
A maintainer who cares about the improvement must measure it.
\end{tcolorbox}

We report the acceptance rate (Section~\ref{sec:rq1:outcome}), the time to a decision (Section~\ref{sec:rq1:process}) and the reasons for rejection, with the re-executed claims (Section~\ref{sec:rq1:grounds}).

\subsection{Acceptance Rate by Agent and Repository Category}
\label{sec:rq1:outcome}

\begin{table}[t]
\caption{Acceptance of the 1,105 closed fixes by agent and by repository category. Categories follow \citet{zhao2023perf} (Section~\ref{sec:design:rq1:approach}); ``other ASF category'' pools the sixteen ASF labels outside their eight, and ``Other'' holds projects with no ASF analogue (games, blockchain, business and scientific applications, AI applications, desktop utilities).}
\label{tab:accept}
\small
\setlength{\tabcolsep}{4pt}
\begin{tabularx}{\textwidth}{>{\raggedright\arraybackslash}Xrrrr}
\toprule
 & Repos & Fixes & Merged & Accepted \\
\midrule
\multicolumn{5}{@{}l}{\emph{By agent}} \\
\rowcolor{gray!12} Codex & 141 & 408 & 296 & 73\% \\
Claude Code & 36 & 51 & 37 & 73\% \\
\rowcolor{gray!12} Jules & 6 & 15 & 11 & 73\% \\
Copilot & 232 & 397 & 200 & 50\% \\
\rowcolor{gray!12} Cursor & 59 & 89 & 37 & 42\% \\
Devin & 57 & 145 & 44 & 30\% \\
\midrule
\multicolumn{5}{@{}l}{\emph{By repository category}} \\
\rowcolor{gray!12} Content & 71 & 116 & 80 & 69\% \\
Library & 49 & 120 & 82 & 68\% \\
\rowcolor{gray!12} Server-client & 37 & 88 & 59 & 67\% \\
Search & 11 & 15 & 9 & 60\% \\
\rowcolor{gray!12} Language & 18 & 143 & 73 & 51\% \\
Cloud & 14 & 20 & 10 & 50\% \\
\rowcolor{gray!12} Big-data & 37 & 66 & 31 & 47\% \\
Build Management & 23 & 50 & 20 & 40\% \\
\rowcolor{gray!12} Other ASF category & 146 & 277 & 151 & 55\% \\
Other & 96 & 210 & 110 & 52\% \\
\midrule
\rowcolor{gray!12} All closed fixes & 502 & 1,105 & 625 & 57\% \\
\bottomrule
\end{tabularx}
\end{table}

\smallskip\noindent\textbf{Overall rate.}
Of the 1,105 closed fixes, \textbf{57\% are merged} (Table~\ref{tab:accept}).
The rate is uneven over time, because in January--April 2025 nearly all fixes were Devin's and acceptance was near zero, whereas since May 2025 the monthly rate has stayed between 53\% and 69\%.
Acceptance also falls with repository size (small 64\%, mid 61\%, large 44\%; \(\chi^2\) = 32.8, V = 0.17) and with the author association that GitHub records for the PR account, since fixes by an OWNER are accepted 77\% of the time, by a CONTRIBUTOR 51\%, and by an author with no association (NONE) 6\% (3/48).

\smallskip\noindent\textbf{By repository category.}
Acceptance differs by the kind of project that received the fix (\(\chi^2\) = 30.2, p \(<\) 0.001, V = 0.17 over the ten categories of Table~\ref{tab:accept}), about as much as it differs by repository size.
Fixes to Content, Library and Server-client projects are accepted in 67--69\% of cases, whereas fixes to Language implementations (51\%), Big-data systems (47\%) and Build Management tools (40\%) are accepted in half or fewer.
The same ordering holds when each repository counts once.
For example, among Content and Library repositories, 56\% and 59\% merged every fix they received, and 35\% and 24\% merged none.
Among Big-data and Build Management repositories, 41\% and 39\% merged every fix, and 49\% and 43\% merged none.
However, two repositories dominate their categories, since mochilang/mochi holds 80 of the 143 closed Language fixes (all by Codex, 54\% accepted) and MihaiCristianCondrea/Smart-Cleaner-for-Android holds 41 of the 277 closed fixes in the other ASF categories (all by Codex, 95\% accepted).
Without these two repositories, the Language rate is 48\%, the other-ASF rate is 47\%, and the category differences remain (V = 0.19).

\smallskip\noindent\textbf{By agent.}
Acceptance differs by agent (\(\chi^2\) = 104, V = 0.31).
Codex, Claude Code and Jules are each accepted in 73\% of cases, Copilot in 50\%, Cursor in 42\% and Devin in 30\%.
The drop with repository size holds within Copilot (61\% \(\rightarrow\) 42\%), is flat within Codex (70\% \(\rightarrow\) 67\%), and reverses for Devin (19\% \(\rightarrow\) 35\%).
The agent ordering largely follows the account model (Section~\ref{sec:population}).
The operator opens Codex, Cursor and Claude Code PRs and merges their own agent's work in 81\%, 95\% and 84\% of merged cases respectively.
In contrast, a bot account opens Copilot, Devin and Jules PRs, and someone else merges them in every case where the merger is recorded.

An operator's self-merge usually takes minutes and leaves no review.
Of the 306 self-merged operator fixes, 209 were merged within an hour, and 97\% of these fast merges carry no human text.
For example, the owner of katspaugh/wavesurfer.js opened a Codex fix (PR 4118) that stores the spectrogram's frequency data so that zooming does not recompute it, and merged the 19-line fix 4 minutes later with no review or comment.
The diff compares a new buffer field with the decoded audio but never assigns that field, so each render from decoded audio recomputes the frequencies.

\begin{table}[t]
\caption{Acceptance by repository category within the four largest agents; --- marks cells with fewer than ten fixes, and the number of fixes is in parentheses.}
\label{tab:agentcat}
\small
\setlength{\tabcolsep}{4pt}
\begin{tabularx}{\textwidth}{>{\raggedright\arraybackslash}Xrrrrr}
\toprule
Category & All & Codex & Copilot & Devin & Cursor \\
\midrule
\rowcolor{gray!12} Content & 69\% (116) & 84\% (44) & 58\% (40) & --- (9) & --- (9) \\
Library & 68\% (120) & 87\% (60) & 68\% (28) & 9\% (11) & 17\% (12) \\
\rowcolor{gray!12} Server-client & 67\% (88) & 80\% (10) & 74\% (50) & 38\% (13) & --- (7) \\
Search & 60\% (15) & --- (5) & --- (4) & --- (2) & --- (0) \\
\rowcolor{gray!12} Language & 51\% (143) & 56\% (91) & 41\% (44) & --- (0) & --- (1) \\
Cloud & 50\% (20) & --- (7) & 42\% (12) & --- (1) & --- (0) \\
\rowcolor{gray!12} Big-data & 47\% (66) & 63\% (30) & 44\% (16) & --- (4) & 33\% (15) \\
Build Management & 40\% (50) & --- (1) & 42\% (31) & 10\% (10) & --- (4) \\
\rowcolor{gray!12} Other ASF category & 55\% (277) & 81\% (101) & 43\% (107) & 29\% (42) & 35\% (20) \\
Other & 52\% (210) & 64\% (59) & 46\% (65) & 42\% (53) & 48\% (21) \\
\midrule
\rowcolor{gray!12} All & 57\% (1,105) & 73\% (408) & 50\% (397) & 30\% (145) & 42\% (89) \\
\bottomrule
\end{tabularx}
\end{table}

\smallskip\noindent\textbf{Agent and category together.}
The agent and category differences overlap, because the agents are unevenly spread over the categories.
The three agents with the highest acceptance open 50\% of the Content fixes and 58\% of the Library fixes but only 10\% of the Build Management fixes and 20\% of the Server-client fixes.
Table~\ref{tab:agentcat} therefore repeats the breakdown inside each agent.
The category ordering holds within the two largest agents.
For example, Codex is accepted in 84--87\% of its Content and Library fixes but in 56\% of its Language fixes and 63\% of its Big-data fixes (\(\chi^2\) = 29.5, p \(<\) 0.001, V = 0.27).
Copilot is accepted in 74\% of its Server-client, 68\% of its Library and 58\% of its Content fixes but in 41--44\% of its Language, Big-data and Build Management fixes (\(\chi^2\) = 21.3, p = 0.011, V = 0.23).
Within each of the two largest agents, maintainers of content and library projects accept the agent's performance fixes more often than maintainers of language implementations, data systems and build tools do.
The agent ordering holds within categories as well.
For instance, within Library projects, acceptance runs from 87\% for Codex to 68\% for Copilot, 17\% for Cursor and 9\% for Devin, and within the pooled other ASF categories from 81\% to 43\%, 35\% and 29\%.
Meanwhile, Devin stays below 45\% in every category with at least ten fixes, and its lowest rate, 10\% in Build Management, comes from ten fixes to nine repositories.
Server-client is the one category where Copilot's rate comes close to Codex's, and one repository carries much of it: frostwire/frostwire merged 15 of its 16 Copilot fixes, and Copilot's rate elsewhere in the category is 65\% (22/34).

\subsection{Time to Acceptance or Rejection}
\label{sec:rq1:process}

\begin{figure}[t]
\centering
\includegraphics[width=0.88\linewidth]{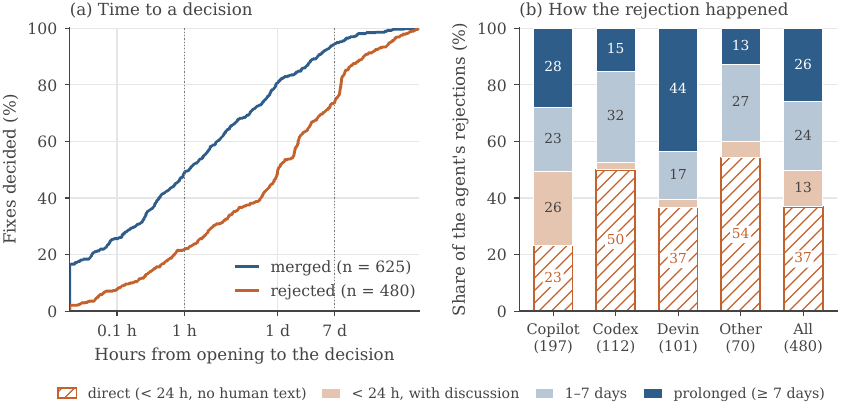}
\caption{Time to a decision: (a) share of merged and rejected fixes decided within a given number of hours of opening; (b) how the 480 rejections happened, by agent (a direct rejection closes within 24 hours with no human text, a prolonged one after seven days or more).}
\label{fig:decision}
\end{figure}
Merged fixes are decided within hours, whereas rejections take about a day and often more than a week (Figure~\ref{fig:decision}a).
Merged fixes close in a median of 1.2 hours (IQR 0.1--16) and rejected ones in 24.3 hours (IQR 1.6--175).
Within an hour, 49\% of merges and 22\% of rejections are decided; 26\% of rejections and 6\% of merges take over a week.

More than a third of the 480 rejections (\textbf{37\%}) are direct rejections, which close within 24 hours with no human text at all (Figure~\ref{fig:decision}b).
Another 26\% of rejections are prolonged.
Repository size shifts this mix, since small repositories reject directly more often than large ones (45\% versus 34\%), while large ones more often reject late (31\% prolonged).
The agents differ more, because Codex and the three smaller agents are rejected directly in half of their cases (50\% and 54\%), whereas Copilot's rejections are the only ones that often carry discussion within the first day (26\%).
Devin's rejections are prolonged in 44\% of cases; 37 close with its inactivity template.

A direct rejection leaves no text from which to recover the reason.
In brahmkshatriya/echo (PR 127), Copilot changed 57 lines of one Kotlin view model so that the lyrics view shows cached lyrics first and loads fresh ones in the background; the repository had received no earlier Copilot PR in AIDev v4, GitHub records the author association as NONE, and the fix was rejected 17 hours later with no review or comment.
The same holds for 21 other direct rejections by Copilot or Devin in repositories that had merged none of the agent's PRs before.

Devin itself closes most of its prolonged rejections, often with no human reply in the thread.
In calcom/cal.com (PR 21556), Devin replaced tRPC calls on the availability page with direct repository calls and wrapped both lookups in \texttt{unstable\_cache} with a one-hour revalidation; no human commented; the PR closed one second after Devin posted ``Closing due to inactivity for more than 7 days'', 209.8 hours after opening, the median duration of the 37 template closures (26 contain no human text).
For such fixes, time to decision reflects Devin's inactivity rule, not a human decision.

\subsection{Reasons for Rejected Performance Issue Fixes}
\label{sec:rq1:grounds}

Most rejections are unexplained, and where we re-execute a rejected fix, its claimed improvement usually does not hold.
We first read the reasons the 480 rejection threads state and then turn to the re-executed sample (Section~\ref{sec:pilot}).

\smallskip\noindent\textbf{Stated reasons.}
Our coding of the 480 rejection threads finds \textbf{no stated reason in 61\%} of them and in 86\% of direct rejections.
In contrast, where a reason is stated, the most common are inactivity (9\%, almost all Devin's template), regression or incorrectness (8\%), a rejected design (6\%), doubt about the performance benefit (6\%) and a superseding change (5\%).
Large repositories state a reason marginally more often than small ones (45\% versus 29\%, p = 0.058).
Only 3\% of rejection threads ask for a measurement, and a human engages substantively in 34\% (0\% of direct rejections).
A rejection thread rarely shows whether anyone checked the claimed improvement.

Where a reviewer questions the benefit, the thread can show the claim failing.
In doodlum/skyrim-community-shaders (PR 1281), Copilot cached the renderer's runtime-data lookups at initialization behind new accessor functions and used them in eight other files.
Within an hour, a reviewer requested changes and asked Copilot to analyze ``what we're saving with this caching approach in real world terms.''
Copilot's reply conceded that ``the performance gains are much smaller than initially described.''
The reviewer then asked the repository owner ``did you have any evidence this was a hot path?'', and the fix closed 4 hours after opening.
This fix is one of the 13 rejections in which a reviewer asks for a measurement.

\smallskip\noindent\textbf{Re-executed claims.}
\label{sec:rq1:validity}
\begin{figure}[t]
\centering
\includegraphics[width=0.84\linewidth]{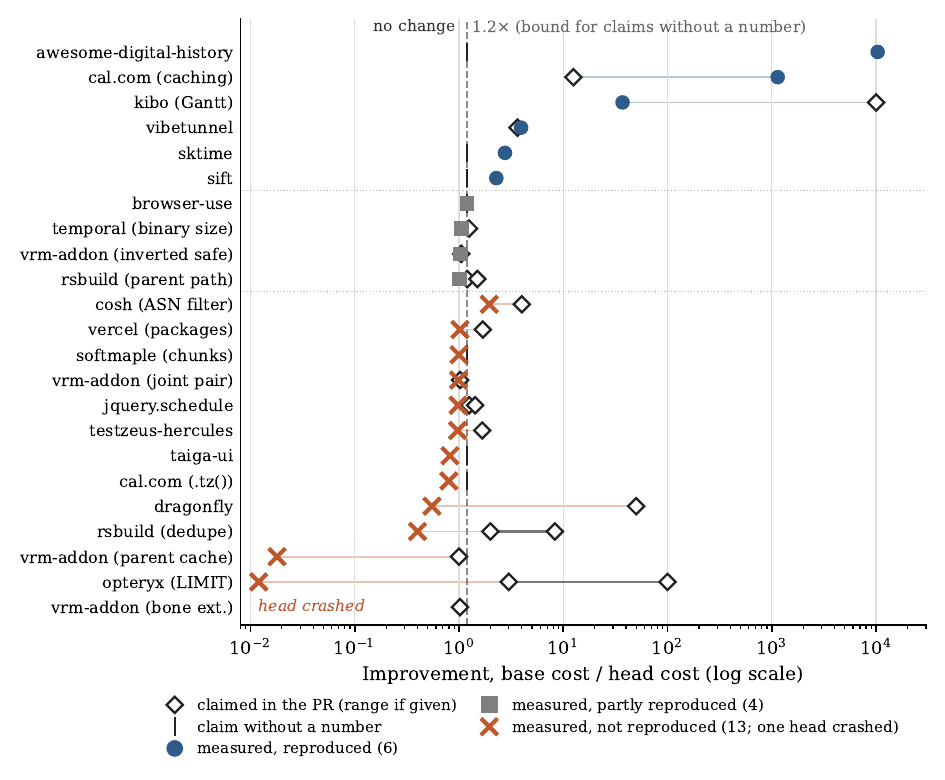}
\caption{Claimed and measured improvement of the 30 rejected fixes drawn for re-execution; the seven that could not run (three need a live database, four are whole-workspace Rust builds) are not shown. Open diamonds mark the claimed improvement (a bar spans a claimed range; a tick at 1.2\(\times\), a claim without a number), filled markers our measurement with the verdict of Eq.~\ref{eq:repro}. A value below 1 is a regression on the workload we ran.}
\label{fig:pilot}
\end{figure}
In the pilot, most rejected fixes with an attached measurement did not deliver the claimed improvement when re-executed, and where a claim gave a number, the measurement fell short of it in all but two of the 16 fixes (Figure~\ref{fig:pilot}).
Of the 23 such fixes, the claim reproduced in 6 (26\%), partly in 4 (17\%), and not at all in 13 (57\%).
Measured on the changed path rather than at production scale (Section~\ref{sec:threats}), the median speed-up is 1.02\(\times\) (IQR 0.86--2.20\(\times\)).
A workload ran on both trees for all 23 fixes, although the base tree built in only 18 cases and the head in 17.
The rest are script projects whose sources run without a build step, so a skipped or failed build did not block the workload.
The tests the fix touched passed on the head in 13 cases (57\%).
However, in several cases the tests the fix added also pass on the base, so they do not exercise the inefficiency they are meant to guard.

The 13 fixes that did not reproduce follow four mechanisms: six run slower, five show no measurable change, one crashes on the head, and one does not contain the claimed change.
First, six fixes run slower.
For example, rsbuild's \texttt{dedupeNestedPaths} claimed 2--7\(\times\) and is slower at every input size (0.40\(\times\) at 1,000 paths), and the opteryx LIMIT push-down moves the limit above the projection and runs 80\(\times\) slower on a 1,000-row query.
A parent-child cache in a Blender add-on makes the many-springs benchmark 57\(\times\) slower, and the PR's own profile used the same branch for both the before and the after run.
The constant-only dragonfly change claims a 98\% CPU reduction and raises the CPU cost per cached query (0.55\(\times\)), the taiga-ui tree that is meant to render only expanded nodes still instantiates all 1,110 of them and renders 18\% slower, and the cal.com \path{.tz()} optimization is 25\% slower and returns half the time slots for one time zone.
Second, five fixes show no change within run-to-run noise (0.97--1.02\(\times\)), and in one of them, softmaple, the chunking the PR body describes is not in the diff, so the build output is byte-identical to the base.
Third, in the same Blender add-on as above, a sibling fix's bone-extension cache segfaults on the head, so its benchmark never completes.
Fourth, the cosh head does not contain the ASN-first reordering the PR text describes, so the 1.95\(\times\) we measured is not the claimed change.

A reproduced speed-up does not by itself show the fix is correct: two of the ten reproduced or partly reproduced fixes break behavior.
The vibetunnel fix cuts WebSocket updates by 75\% on the PR's bursty stream, but on a steady stream its debounce never fires, the terminal animation stalls for four seconds, and the head fails an existing buffer test that the base passes.
The browser-use fix is 1.19\(\times\) faster but mislabels computed styles and finds 1,334 of the 2,000 interactive elements on our test page.

By contrast, the other reproduced fixes deliver concrete gains on the path each fix targets.
In kibo, the Gantt re-render count falls 37\(\times\).
In sift, allocations fall 8\(\times\) (218 \(\rightarrow\) 27 per op).
In sktime, the serialized size shrinks 2.8\(\times\).
In awesome-digital-history, the fix removes a page-load loop of 20,000 pushState calls in ten seconds.

Whether any maintainer checked a claim before rejecting it therefore remains unknown, because most rejections state no reason, whereas the pilot shows that such a check would have failed for most of the re-executed claims.

\section{RQ2: What factors are associated with the acceptance or rejection of agents' performance fixes?}
\label{sec:rq2}

\begin{tcolorbox}[enhanced, colback=gray!10, colframe=black!75, boxrule=0.8pt, arc=3pt, drop shadow={black!50!white}, left=5pt, right=5pt, top=3pt, bottom=3pt]
\textbf{Insight:} The strongest correlate of whether an agent's performance fix is merged is the repository's prior treatment of agent PRs.
Among fix properties, shape is the one that holds up within agent and repository: fixes that remove code are merged more often than fixes that add machinery.
Studies that use the merge as a quality label should control for the repository's history with the agent.
\end{tcolorbox}

The repository's history with agent PRs and the deleted-line share of the fix distinguish merged from rejected fixes, while the coded content of the fix, its description, its tests and its attached measurements do not.
Section~\ref{sec:rq2:pooled} compares the two outcomes in the pooled data and fits the nested models of Section~\ref{sec:design:rq2:metrics}; Sections~\ref{sec:rq2:within} to~\ref{sec:rq2:context} then take up the three variables that the models single out: what survives once the agent is held fixed, whether the patch adds or removes code, and the repository's history with agent PRs.
These are associations in observational data, not measured causes of any decision, and a null result here means that we did not detect a difference, not that none exists; we report the estimates and their intervals so that the reader can see how large a difference the data leave room for.

\subsection{Overall Comparison of Merged and Rejected Fixes}
\label{sec:rq2:pooled}
\label{sec:rq2:models}

The repository-history block raises the fit of the acceptance model most, the deleted-line share is the one content variable with a stable effect, and the coded flags for added machinery add nothing.
The control variables alone explain little because merged and rejected fixes look alike in the coded content.
Root cause and performance category are distributed alike across the outcomes (p = 0.44 and 0.25), and test co-change is level (35\% among merged and 37\% among rejected fixes, p = 0.49).
An attached measurement (20\% versus 15\% for rejected against merged, p = 0.024) and a numeric claim in the text (35\% versus 28\%) are slightly more common among rejected fixes; the lean follows the agent mix, and the measurement's coefficient is not significant once the agent is held fixed (M4).
Merged fixes are smaller (median 90 versus 159 changed lines, p \textless{} 0.001) and more often functional-level (48\% versus 39\%).
A third-party review (Section~\ref{sec:funnel}) is equally common (40\% versus 38\%), and a third-party approval is the one process variable that separates the outcomes (28\% versus 1\%).
The control model M0 of Table~\ref{tab:models} reaches a pseudo R\(^2\) of 0.053, and in it only size, third-party review and repository stars are significant; the review effect describes the path to a merge rather than a property of the submission, since review happens after a PR opens.

\begin{table}[t]
\caption{Nested logistic models of acceptance (n = 1,102 closed fixes with file data; standard errors clustered by repository). M1--M4 add one factor block each to M0; the block test is a likelihood-ratio test against the previous model. Adding the same-agent track record to M4 raises the pseudo R\(^2\) to 0.246, but the track record is collinear with the pre-opening rate.}
\label{tab:models}
\small
\setlength{\tabcolsep}{4pt}
\begin{tabularx}{\textwidth}{l>{\raggedright\arraybackslash}Xrrr}
\toprule
Model & Variables & Pseudo R$^2$ & Block LR $\chi^2$ (df) & p \\
\midrule
\rowcolor{gray!12} M0 & controls: root cause, attached measurement, locality, test change, size, third-party review, stars & 0.053 & & \\
M1 & + F1: agent, description style & 0.119 & 99.2 (9) & \textless{} 0.001 \\
\rowcolor{gray!12} M2 & + F2: deleted-line share, files, noise files, churn share of largest file & 0.144 & 37.3 (4) & \textless{} 0.001 \\
M3 & + F3: adds production file, architectural-level, new shared abstraction & 0.145 & 1.9 (3) & 0.59 \\
\rowcolor{gray!12} M4 & + F4: author association, pre-opening merge rate, agent PRs in repository & 0.232 & 130.9 (8) & \textless{} 0.001 \\
\bottomrule
\end{tabularx}
\end{table}

The four blocks contribute unequally.
F1 lifts the pseudo R\(^2\) from 0.053 to 0.119, but the lift is the agent's identity rather than its prose: the controls with the agent alone reach 0.106, so the six description features add 0.013.
F2 lifts it to 0.144, and the deleted-line share is the block's only significant term (OR 7.1 per unit of share in M2, 95\% CI 2.9--17.0).
F3 adds 0.001 (likelihood-ratio p = 0.59), and none of the three coded forms of added machinery is significant in M3 or in the full model; Section~\ref{sec:rq2:shape} shows that the flags fail to measure the property, whereas the property itself is present.
F4 lifts it from 0.145 to 0.232, and dropping one block at a time from the full model M4 gives the same order (0.145 without F4, 0.210 without F2 and 0.217 without F1); adding the same-agent track record on top raises the fit to 0.246 but splits the rate's coefficient, because the two measure the same history.
In M4, the pre-opening merge rate (OR 13.7 from 0 to 1, 95\% CI 6.3--30.0), an author association of NONE (OR 0.10), MEMBER (2.2) or OWNER (3.0), the deleted-line share (OR 10.5, 95\% CI 4.4--25.2), third-party review, the agent's footer, size and Devin against Copilot remain significant, whereas root cause, the attached measurement, locality, a test change, stars, every other description feature and every F3 term are not.
The stars effect of the controls (OR 0.58 per log unit in M0) disappears once the merge rate enters (OR 0.98, p = 0.90), so a repository's popularity stands in for how it treats agent PRs.

\subsection{Comparison Within Each Agent}
\label{sec:rq2:within}

\begin{figure}[t]
\centering
\includegraphics[width=0.88\linewidth]{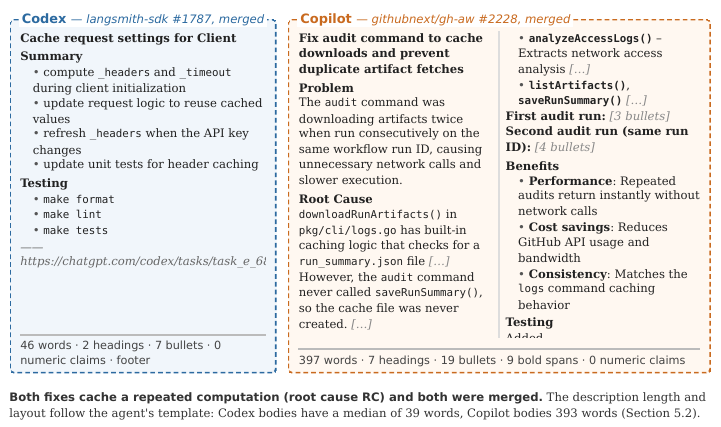}
\caption{Two descriptions of merged fixes that cache a repeated computation, one by Codex and one by Copilot, condensed from the PR bodies. The length and layout follow the agent's template rather than the outcome (Section~\ref{sec:rq2:within}).}
\label{fig:desc}
\end{figure}
Verbosity and ``AI flavor'' differ by outcome only in the pooled data, because they mark each agent's house style (Figure~\ref{fig:desc}).
Pooled, rejected fixes are wordier (median 271 versus 97 words) and carry more headings, bullets, bold text, emoji and hype words (all p \textless{} 0.01).
Within Copilot, Codex and Devin separately, however, description length (p = 0.75, 0.89 and 0.53), bullets, bold text, hype rate and numeric claims do not differ; the only differences are one extra heading and a ``This PR \ldots{}'' opening among rejected Copilot fixes, and across the 74 repositories that hold both outcomes the per-repository differences are zero.
The pooled gap is therefore explained by the agents' templates: Codex writes bodies of about 40 words and is accepted most, while Copilot and Devin write 250--400-word bodies and are accepted least.
This is why the F1 block of Table~\ref{tab:models} is the agent and not the prose, and in the full model only the agent's footer and Devin remain significant among the F1 terms.
The footer is the generation note that Codex and Devin append by default (82\% and 93\% of their fixes) and Copilot almost never does (2\%), so its coefficient is a within-agent contrast among their fixes with and without the note, not a property of what the description says.

The repository category of Section~\ref{sec:rq1:outcome} does not change this reading.
Within a category, rejected fixes are wordier only where the two outcomes hold different agents: in Library projects, merged fixes have a median of 45 words and rejected ones 217 (p = 0.006), and Codex wrote 63\% of the merged fixes but 21\% of the rejected ones, and the same holds in the other ASF categories and in the Other group.
Within category and agent together, the twelve cells with at least ten fixes per outcome show no difference in length, apart from two cells whose medians differ by 9 and 51 words (p = 0.034 and 0.040).
What a category's maintainers read is therefore the agent's template, and the outcome does not follow it within any category.

\subsection{Whether the Patch Adds or Removes Code}
\label{sec:rq2:shape}
\label{sec:rq2:pairs}

\begin{figure}[t]
\centering
\includegraphics[width=0.62\linewidth]{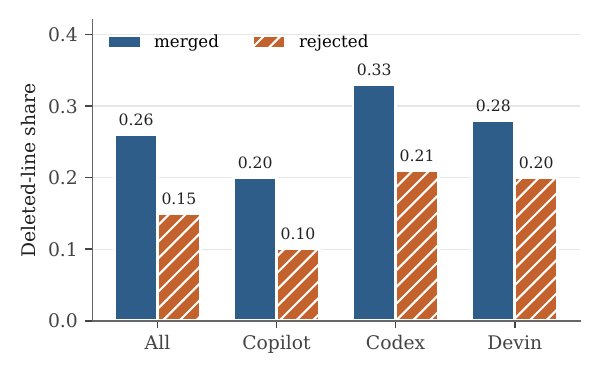}
\caption{Deleted-line share of merged and rejected fixes, pooled and within the three largest agents.}
\label{fig:delshare}
\end{figure}
Merged fixes have a higher deleted-line share than rejected fixes in the pooled data, within each agent and within repositories (Figure~\ref{fig:delshare}).
The share (Eq.~\ref{eq:delshare}) is 0.26 among merged and 0.15 among rejected fixes pooled, and 0.20 versus 0.10 within Copilot, 0.33 versus 0.21 within Codex and 0.28 versus 0.20 within Devin (all p \(\le\) 0.001); merged fixes delete a larger share in 50 of the 74 repositories that hold both outcomes, a smaller one in 23 and the same share in one (Wilcoxon p \textless{} 0.001).
In the models the share enters with OR 7.1 in M2 and 10.5 in the full model M4, that is, the odds of acceptance rise by a factor of 1.3 for every ten points of share with the agent, the description, the coded content and the repository's history held fixed.
Rejected fixes also add new production files more often (24\% versus 15\%, p \textless{} 0.001), but no other shape feature is as consistent: diff size differs within Copilot (196 versus 110 lines, p = 0.001) but not within Codex or Devin, and file count, directories, lock files, generated files and documentation differ in no comparison.

\begin{figure}[t]
\centering
\includegraphics[width=0.88\linewidth]{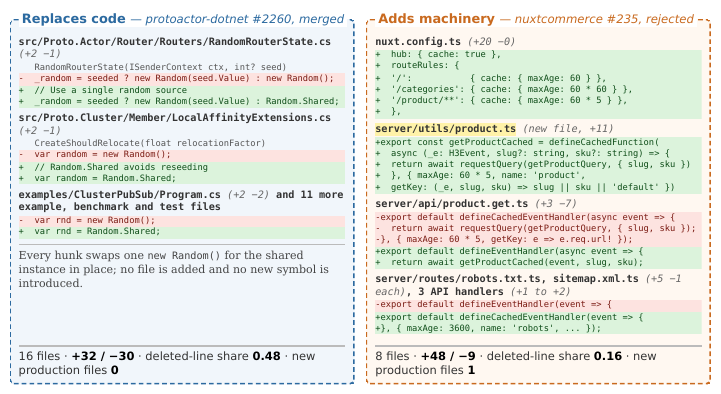}
\caption{Two Codex fixes of similar size that replace code (left) and add machinery (right), condensed from their diffs. The merged fix swaps one call at each site; the rejected fix adds cache rules, a new production file and a cached wrapper around the existing handlers.}
\label{fig:shape}
\end{figure}
Two Codex fixes of similar size show the difference between replacing code and adding it (Figure~\ref{fig:shape}).
In asynkron/protoactor-dotnet (PR 2260), the fix replaces each per-call \texttt{new Random()} with the shared \texttt{Random.Shared} instance, adding 32 lines and deleting 30 across 16 files; its deleted-line share is 0.48, and the operator merged it eight minutes after opening it.
The zackha/nuxtcommerce fix (PR 235) enables NuxtHub caching with 48 added and 9 deleted lines across eight files, writing cache rules into \texttt{nuxt.config.ts} and adding the new production file \texttt{server/utils/product.ts}, which wraps the product query in a cached function; its share is 0.16, and the repository owner who opened it through Codex closed it after four days with an empty thread.
The coders classed both fixes as architectural-level optimization clones.

The blind pairs test the same distinction with the repository and the agent held fixed by construction, and show that a reader sees more of it than the deleted-line share measures.
Given one merged and one rejected fix from the same repository and agent, a blind judge named the merged fix as the one it would merge (its would-merge judgment, Section~\ref{sec:design:rq2:approach}) in 33 of 40 pairs (binomial p = 4 \(\times\) \(10^{-5}\)), more often than any measured feature does.
Its reasons point to readability, calibrated claims and the difference between replacing code and adding it: it named the rejected fix as the one that adds more machinery (new caches, options, helper modules) in 25 of the 40 pairs and the merged one in 9, classified the change style of the merged fix as \emph{replaces} 26 times and \emph{adds} 5 times against 13 and 19 for the rejected fix, found the merged fix easier to read and trust in 31 of the 40 pairs (reading cost 2.4 versus 3.3 on a 1--5 scale), and rated the rejected description as overclaiming twice as often (14 versus 7).

The judge's choice agrees with the outcome more often than any single measured feature does.
Within the same 40 pairs, the fix with the higher deleted-line share is the merged one in only 24, the smaller fix in 23, and the coded flags of the F3 block are balanced (the merged fix adds a production file in 8 pairs and the rejected one in 10, and the coders judged 20 merged and 19 rejected fixes architectural-level).
The judge is nevertheless right in 12 of the 16 pairs in which the share points the wrong way, and what it calls added machinery overlaps with the share without coinciding with it: in the 25 pairs in which it names the rejected fix as the one that adds machinery, the merged fix's share exceeds the rejected fix's by a median of 0.22, and in the other 15 pairs by 0.01.
This is why the F3 block leaves the pseudo R\(^2\) unchanged in Table~\ref{tab:models}: its three flags are binary and broad (an architectural-level scope covers 45\% of the closed fixes and a production file is added by 19\%), they differ modestly by outcome in the pooled data (24\% versus 15\%, 49\% versus 43\% and 16\% versus 14\%), and their odds ratios in M3 are 0.84, 0.97 and 1.36 (all p \textgreater{} 0.28), with or without the share in the model.
The property that the judge reads off two concrete diffs, whether the fix installs a mechanism that the problem did not require, is therefore detectable in these same-repository, same-agent pairs, but our coded proxies do not measure it and the deleted-line share measures only part of it.
The judge is a language model and the pairs are few, so this is evidence that the difference is visible to a careful reader, not a measurement of how often maintainers read that carefully.

A Copilot pair from doodlum/skyrim-community-shaders shows what the judge counted as added machinery.
The merged fix (PR 1368) adds 7 lines and deletes 6 in two files to run the overlay drag update only when an overlay is visible (share 0.46), and was merged after 2.4 hours.
The rejected fix (PR 1281) adds 123 lines and deletes 40 across ten files (share 0.25), almost half of them in \texttt{Globals.h} and \texttt{Globals.cpp}, where it declares cached pointers to the game's runtime data and accessors that read them (Section~\ref{sec:rq1:grounds} gives the reviewer exchange that closed it).
The judge picked the merged fix and wrote that the rejected one ``adds global void* caches with unchecked casts, claims 50+ calls saved per frame without measurement, and risks stale pointers.''

\subsection{Repository History with Agent PRs}
\label{sec:rq2:context}

Acceptance differs more across the repository's history with agent PRs, read off the pre-opening merge rate (Eq.~\ref{eq:mr}), than across any measured property of the fix, and the deleted-line share carries no signal where the repository rarely merges agent PRs and a strong one where it decides case by case.
\begin{figure}[t]
\centering
\includegraphics[width=0.88\linewidth]{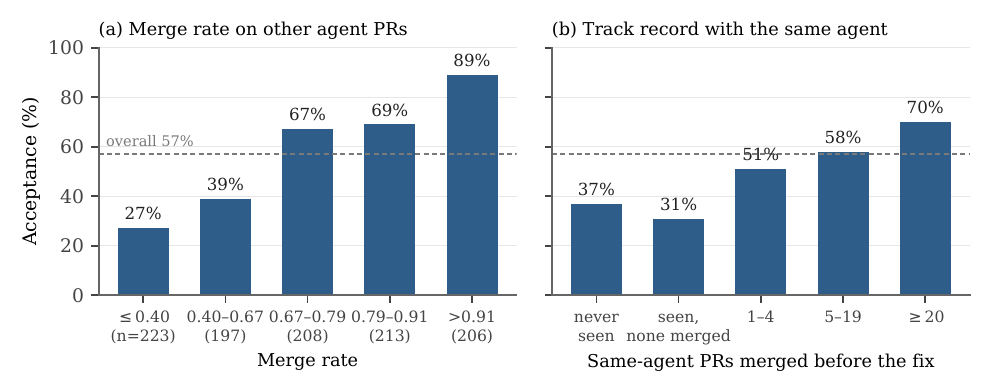}
\caption{Acceptance of agents' performance fixes by (a) the repository's snapshot merge rate on its other agent PRs and (b) the number of the same agent's PRs the repository had merged before the fix opened. The dashed line is the overall acceptance of 57\%.}
\label{fig:reporate}
\end{figure}
Acceptance is 33\% where the repository had merged at most 40\% of its decided agent PRs when the fix opened, 57\% in the middle band and 84\% where it had merged over 90\% (n = 165, 521 and 255; \(\chi^2\) = 109.7, p \textless{} 0.001), and the pre-opening quintiles run 33, 43, 70, 74 and 80\%.
In the full model, moving this rate from 0 to 1 multiplies the odds of acceptance by 13.7 (95\% CI 6.3--30.0), that is, by 1.30 per ten percentage points, and the F4 block lifts the pseudo R\(^2\) from 0.145 to 0.232 (M4).
Among the 755 fixes whose repositories had at least five decided agent PRs before opening, the same coefficient is larger still (OR 61, 95\% CI 16--235).

\begin{table}[t]
\caption{Acceptance by the repository's snapshot merge rate on its other agent PRs (over all of AIDev v4; the models use the pre-opening rate of Eq.~\ref{eq:mr}), pooled and within the three largest agents. The bins are quintiles of the snapshot rate (edges 0.399, 0.665, 0.793, 0.909) over the 1,047 modeled fixes whose repositories have other agent PRs; --- marks cells with fewer than ten fixes.}
\label{tab:reporate}
\small
\setlength{\tabcolsep}{4pt}
\begin{tabularx}{\textwidth}{>{\raggedright\arraybackslash}Xrrrrr}
\toprule
Snapshot merge rate on other agent PRs & n & Accepted & Copilot & Codex & Devin \\
\midrule
\rowcolor{gray!12} $\le$ 0.40 & 223 & 27\% & 37\% & 32\% & 18\% \\
0.40--0.67 & 197 & 39\% & 32\% & 53\% & 40\% \\
\rowcolor{gray!12} 0.67--0.79 & 208 & 67\% & 69\% & 63\% & 53\% \\
0.79--0.91 & 213 & 69\% & 64\% & 68\% & --- \\
\rowcolor{gray!12} $>$ 0.91 & 206 & 89\% & 79\% & 96\% & --- \\
\bottomrule
\end{tabularx}
\end{table}

Table~\ref{tab:reporate} and Figure~\ref{fig:reporate}a report the snapshot variant of the rate, computed over all of AIDev v4 and so also over decisions made after the fix opened.
It orders the bands the same way and fits better (pseudo R\(^2\) 0.253; OR 32.6), precisely because later decisions carry information no submitter could have had, which is why the pre-opening rate carries the interpretation.
The snapshot bands describe three kinds of repository.
At the low end are repositories that do not take agent PRs: the 47 fixes with file data whose author association is NONE are rejected 94\% of the time, and 42 of them come from Copilot and Devin bot accounts.
At the high end are repositories that take nearly all of them: in the band over 0.91, Codex fixes are accepted 96\% of the time, and 97\% of these merged fixes received no third-party review.
Between the two lie the repositories that decide case by case.
Within the 521 middle-band fixes, the deleted-line share separates merged from rejected fixes (OR 8.6 per unit of share with size held fixed, p \textless{} 0.001); within the 165 low-band fixes it does not (OR 1.5--2.1 across specifications, p \(\ge\) 0.23), and in the high band the estimate is unstable across specifications.
The share therefore carries no signal where the repository rarely merges agent PRs and a strong one where it decides case by case, and whether it thins out again at the high end the data do not settle.

The same-agent track record (Eq.~\ref{eq:tr}) shows the same gradient (Section~\ref{sec:threats} discusses its limits).
Acceptance is 37\% when the repository had never seen the agent (Figure~\ref{fig:reporate}b), 31\% when it had seen the agent but merged nothing, 51\% after 1--4 merged PRs, 58\% after 5--19 and 70\% after 20 or more.

Silent decisions remove one mediator from the comparison, because no review text stands between the submission and the outcome.
Among the Copilot fixes closed with no human text (41 silently merged, 70 silently rejected), description length, headings, claims and diff size are the same on both sides, but the silently merged fixes come from repositories with a higher merge rate (0.77 versus 0.45, p \textless{} 0.001) and a longer Copilot track record (13 versus 1.5 merged PRs) and have a higher deleted-line share (0.23 versus 0.13).
For Codex (264 versus 101), the merge rate (0.90 versus 0.83, p \textless{} 0.001) and the deleted-line share (0.32 versus 0.21, p = 0.002) differ; for Devin (10 versus 72), the merge rate differs (0.80 versus 0.39) while the share does not (0.23 versus 0.20, p = 0.45).

Two silent Copilot fixes with descriptions of the same length and layout (477 and 480 words, 7 headings and 15 bullets each) landed in repositories with different histories.
In jirihofman/portfolio (PR 279), Copilot replaced a REST query with a GraphQL count query in two files, and the owner merged the fix 18 minutes later without a comment; in bryanedds/Nu (PR 1074), Copilot split large terrains into patches with their own geometry in three files, and the PR closed after seven days with an empty thread.
The first repository had merged 7 of its 8 other closed agent PRs and five Copilot PRs, whereas the second had merged 2 of 14 and none from Copilot, and the merged fix also has the higher deleted-line share (0.44 versus 0.09).

\section{RQ3: How frequently do agents' performance fixes involve architectural changes?}
\label{sec:rq3}

\begin{tcolorbox}[enhanced, colback=gray!10, colframe=black!75, boxrule=0.8pt, arc=3pt, drop shadow={black!50!white}, left=5pt, right=5pt, top=3pt, bottom=3pt]
\textbf{Insight:} Agents fix the same inefficiencies as human developers and resolve them most often by caching.
However, they change several files for one mechanism more often than the human fixes \citet{zhao2023perf} coded, with a small edit in each file.
Reach reflects how an agent optimizes, not what it optimizes.
\end{tcolorbox}

The performance issues that agents fix concentrate on repeated computation, and the fixes most often resolve it by caching.
The fixes are architectural-level in 46\% of cases, change one or two functions per file, and their heaviest pattern, the new shared abstraction, adds the most code.
We first describe the issues by root cause and the resolution agents apply to each (Section~\ref{sec:rq3:causes}), then separate functional-level from architectural-level fixes (Section~\ref{sec:rq3:scope}), and finally measure the reach of architectural-level fixes and relate it to the deleted-line share of RQ2 (Section~\ref{sec:rq3:depth}).

\subsection{Root Causes and How Agents Fix Them}
\label{sec:rq3:causes}

Repeated computation (RC, 27\%) and redundant data processing (RDP, 17\%) together account for 44\% of the performance issues that agents fix.
The remaining issues spread over general inefficient computation (GIC, 16\%), inefficiency under special cases (ISC, 11\%), inefficient API usage (IAU, 10\%), synchronization (7\%), data structures (7\%) and iteration (3\%) (Table~\ref{tab:rootcause}).
The profile differs by agent (V = 0.16).
Repeated computation accounts for 35\% of the issues that Codex and Devin fix, whereas Copilot's issues spread more over ISC and IAU (15\% each).
However, root cause does not predict acceptance (RC 59\%, RDP 56\%, GIC 50\%, ISC 56\%; p = 0.44).

\begin{table}[t]
\caption{Root causes of the performance issues agents fix (consensus set), pooled and by agent. Codes follow \citet{zhao2023perf}; Other pools Cursor, Claude Code and Jules.}
\label{tab:rootcause}
\small
\setlength{\tabcolsep}{4pt}
\begin{tabularx}{\textwidth}{>{\raggedright\arraybackslash}Xrrrrr}
\toprule
 & All & Copilot & Codex & Devin & Other \\
Root cause & (1,262) & (488) & (438) & (147) & (189) \\
\midrule
\rowcolor{gray!12} RC repeated computation & 27\% & 19\% & 35\% & 35\% & 25\% \\
RDP redundant data processing & 17\% & 16\% & 13\% & 24\% & 21\% \\
\rowcolor{gray!12} GIC general inefficient computation & 16\% & 13\% & 20\% & 13\% & 16\% \\
ISC inefficiency under special cases & 11\% & 15\% & 8\% & 6\% & 13\% \\
\rowcolor{gray!12} IAU inefficient API usage & 10\% & 15\% & 8\% & 7\% & 4\% \\
IS inefficient synchronization & 7\% & 8\% & 6\% & 4\% & 7\% \\
\rowcolor{gray!12} IDS inefficient data structure & 7\% & 9\% & 5\% & 5\% & 9\% \\
II inefficient iteration & 3\% & 3\% & 3\% & 3\% & 1\% \\
\rowcolor{gray!12} unclear & 2\% & 1\% & 1\% & 1\% & 4\% \\
\bottomrule
\end{tabularx}
\end{table}

Agents most often remove the inefficiency by caching or memoization.
Under our fifteen-code book, caching or memoization is the primary resolution of 20\% of fixes, ahead of concurrency change (10\%), removing redundant work (8\%), API replacement (8\%) and resource release or bounding (7\%).
Early exit or fast path, batching or coalescing, lazy or deferred evaluation, allocation reduction and algorithm change account for 6\% each, data-structure change and I/O or query tuning for 5\% each, and the remaining codes for \(\le\) 2\% each.
Codex and Devin resolve 26\% of fixes by caching, Copilot 14\% (V = 0.18); Copilot more often replaces APIs and bounds resources.

Each root cause except redundant data processing has one dominant resolution.
Fixes of repeated computation are resolved by caching in 67\% of cases, and fixes of inefficient synchronization by a concurrency change in 84\%.
Similarly, inefficient API usage is resolved by API replacement in 63\% of cases, and inefficient data structures by a data-structure change in 57\%.
Inefficiency under special cases, which in the agent fixes is mostly a leak or an unbounded resource, is resolved by resource release or bounding in 52\% of cases.
Redundant data processing is resolved by batching in 23\% and by redundant-work removal, lazy evaluation and query tuning in 17\% each.

Resolution and outcome are unrelated overall (p = 0.11), with one exception.
Lazy or deferred evaluation is accepted 40\% of the time against 57\% overall, and it appears in 10\% of rejected and 5\% of merged fixes (p = 0.004).
At the other end, allocation reduction is accepted most often (69\%).
\citet{msr2026perfopt} found acceptance varying with the type of optimization in pooled agent PRs; under our hand-coded resolutions the difference does not survive except for lazy evaluation, and Section~\ref{sec:rq2:within} shows how a pooled difference can reflect the agent mix.

\enlargethispage{10pt}%
The fixes at the two ends of this range differ in whether they remove work or move it to a later point.
In plengauer/Thoth (PR 2255), Copilot deferred the heavy OpenTelemetry imports of \path{sdk.py} from module load to the first INIT command, and the PR description reports module load falling from about 170 ms to 13 ms.
However, the maintainer replied that ``there is always an INIT'', so the deferred imports would still run in every session; he judged the change not worth it on that ground, and the PR was closed 2.2 hours after it opened.
In contrast, in dotnet/aspnetcore (PR 62056) Copilot replaced two arrays that the validation filter allocated per endpoint parameter with a list of only the parameters that need validation.
The change stays in one file (21 lines added, 16 deleted).
The maintainer noted that the filter has no benchmark and judged that ``assuming the change doesn't break any tests the delta here feels straightforward to review.''
It was merged after 6.8 days.

\subsection{Functional-level and Architectural-level Fixes}
\label{sec:rq3:scope}

Coded by mechanism, 46\% of all agent fixes are architectural-level (582 of 1,259).\footnote{That this count equals the 582 repositories of the consensus set is a coincidence.}
Of the 721 multi-file consensus fixes, the coders judged 81\% architectural-level, 11\% a functional-level change plus noise (lock files, formatting, unrelated edits), and 7\% a performance change bundled with unrelated work.
Noise affects 26\% of two-file fixes against 5\% of fixes with three or more files, so it concentrates in the smallest multi-file fixes.
Architectural-level fixes are accepted 53\% of the time against 60\% for the rest (p = 0.044); tests, measurements and third-party review do not differ.

Generated build output is one of the noise kinds that make a functional-level fix look multi-file.
In ateliee/jquery.schedule (PR 58), Devin cached repeated jQuery DOM lookups in one source file but also committed the rebuilt script, its minified copy and its source map under \path{dist/}, so a file count sees four production files where the coders judged one relevant.
Devin's account closed it after seven days of inactivity.

A raw file count would put the architectural-level share at 57\%, since it also counts multi-file fixes whose extra files are noise or unrelated work.
Of the 1,259 fixes with file data, 43\% touch at most one production file, 17\% touch two and 41\% three or more (rounding makes these shares sum to 101\%).
Multi-file fixes change more code (median 148 versus 30 production lines) and functions (7 versus 2) than functional-level ones and more often change tests (40\% versus 34\%) and attach measurements (20\% versus 15\%).
However, acceptance falls as a fix touches more files, from 62\% for functional-level fixes to 57\% for two files and 51\% for three or more (p = 0.004).
By agent, Codex is functional-level in 57\% of fixes, Copilot and Devin in 35--38\%.

\begin{figure}[t]
\centering
\includegraphics[width=0.85\linewidth]{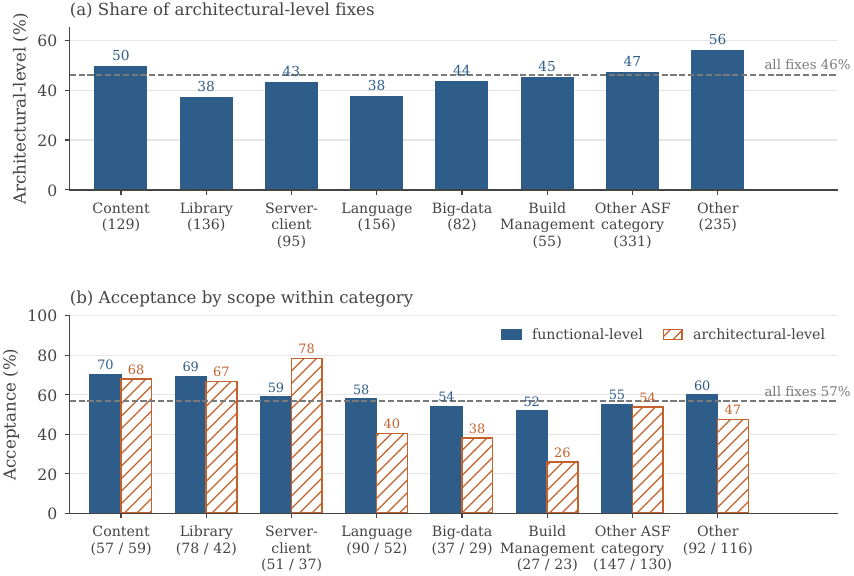}
\caption{Architectural-level fixes by repository category (Section~\ref{sec:rq1:outcome}; Search and Cloud omitted for fewer than ten fixes per scope): (a) share of fixes coded architectural-level; (b) acceptance of each scope within each category; numbers under the labels are the fixes in (a) and the closed fixes per scope in (b).}
\label{fig:archcat}
\end{figure}
By repository category, the architectural-level share runs from 38\% in Library and Language projects to 56\% in the Other group (\(\chi^2\) = 19.6, p = 0.021, V = 0.12; Figure~\ref{fig:archcat}a), and without the two dominant repositories of Section~\ref{sec:rq1:outcome} the Language share rises to 63\% (V = 0.14).
The acceptance penalty of architectural-level fixes, however, is concentrated in the categories that accept the fewest fixes (Figure~\ref{fig:archcat}b).
In Content, Library and Server-client projects, which accept 67--69\% of fixes, the two scopes are accepted alike (70\% versus 68\% and 69\% versus 67\%) or the architectural-level fixes more often (59\% versus 78\% in Server-client), and over the three categories the rates are 67\% and 70\% (p = 0.55).
In Language, Big-data and Build Management projects, which accept 40--51\%, architectural-level fixes are accepted at 40\%, 38\% and 26\% against 58\%, 54\% and 52\% for functional-level ones, that is, 37\% against 56\% over the three categories (p = 0.003).
The difference holds within the two largest agents, since Codex's architectural-level fixes are accepted at 46\% against 62\% for its functional-level fixes in these three categories and at 86\% against 85\% in the first three, and Copilot's at 35\% against 51\% and at 71\% against 63\%.
The maintainers of language implementations, data systems and build tools therefore accept a fix that reaches several files less often, whereas the maintainers of content, library and server-client projects do not distinguish the two scopes.

Agents add a new shared abstraction as often as they propagate a changed interface.
Among the architectural-level fixes with GitHub patches (Table~\ref{tab:patterns}), a changed interface propagated to its callers and a new shared abstraction (cache, pool, helper) used from several places each account for 31\%, the same optimization cloned into several sites for 23\%, several independent optimizations in one fix for 14\%, and a classic design pattern for 1\%.
However, acceptance does not differ across patterns (47--59\%, p = 0.48).

\subsection{Reach and Acceptance of Architectural-level Fixes}
\label{sec:rq3:depth}

An architectural-level fix makes a small edit in each of several files, touching a median of 7 functions (IQR 4--13) across 4 files (2--7) in 9 hunks and 151 production lines.
By this count (Section~\ref{sec:design:rq3:metrics}), 72\% of architectural-level fixes touch five or more functions and 37\% ten or more, against 11\% and 1\% of functional-level fixes.
The edit inside each file, however, is as small as in a functional-level fix: 1.7 functions per file against 2.0 (p = 0.42), 18 lines per function against 17.
By agent, Copilot's architectural-level fixes touch 8 functions and 166 lines, Codex's 6 and 99, and Devin's 6 and 130.
By outcome, merged and rejected architectural-level fixes touch the same number of files (4 versus 4) and functions (8 versus 7, p = 0.84) and differ only in lines (128 versus 186, p = 0.025).

The four main architectural-level patterns touch a similar number of functions (Kruskal--Wallis p = 0.31) but differ in size and shape, and the \emph{new shared abstraction} is the heaviest of them (Table~\ref{tab:patterns}; lines there are production lines changed).
It changes 258 production lines at 28 lines per function, has the lowest deleted-line share (0.10), and adds a new production file in 63\% of cases.
In contrast, the \emph{optimization clone} is the lightest, with 10 lines per function and the highest deleted-line share (0.40).

\begin{figure}[t]
\centering
\includegraphics[width=0.88\linewidth]{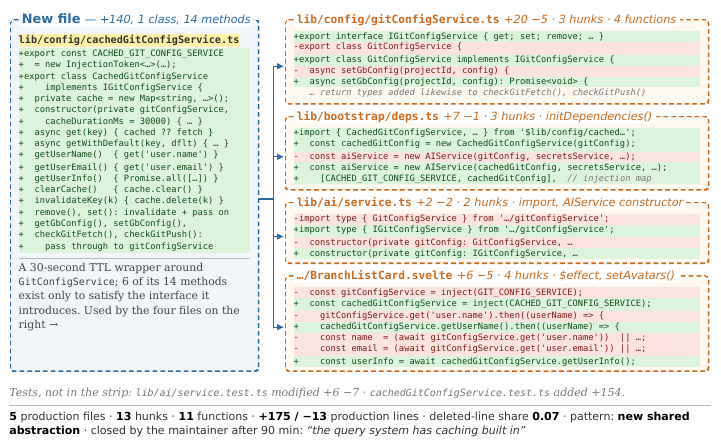}
\caption{Reach of one architectural-level fix (gitbutlerapp/gitbutler PR 10222, Copilot, new shared abstraction), condensed from its diff. The new cache service on the left is used from the four production files on the right, each changed in one to four hunks of a few lines; the fix touches 11 functions in 5 files and deletes 13 of its 188 production lines.}
\label{fig:reach}
\end{figure}
A cache serving several call sites shows why the shared abstraction is the heaviest pattern (Figure~\ref{fig:reach}).
In gitbutlerapp/gitbutler (PR 10222), each branch card read the user's name and email from the global git configuration with two calls per branch.
Copilot added a new \texttt{CachedGitConfigService} that keeps these values for 30 seconds and used it from \texttt{BranchListCard} and \texttt{AIService}.
The fix changes 188 lines in five production files, and only 13 of them are deletions (deleted-line share 0.07).
However, the maintainer closed it within 90 minutes and wrote that ``the query system has caching built in and maybe it can cache this while on the page''.
In contrast, in assistant-ui/assistant-ui (PR 2113) Devin applied the same caching resolution at five sites by wrapping the \texttt{Array.from} call of five React render functions in \texttt{useMemo}.
This optimization clone changes 101 production lines at one function per file, deletes 45 of them (0.45), and was merged 19 minutes after it opened with no human comment.
Both fixes cache a repeated computation, yet the gitbutler fix adds a service beside the repository's caching layer, whereas the assistant-ui fix rewrites existing lines in place.

\begin{table}[t]
\caption{Reach of architectural-level fixes by pattern (medians; consensus fixes with GitHub patches). Del.\ share is the deleted-line share; New file is the share adding a production file.}
\label{tab:patterns}
\small
\setlength{\tabcolsep}{4pt}
\begin{tabularx}{\textwidth}{>{\raggedright\arraybackslash}Xrrrrrrr}
\toprule
Pattern & n & Files & Lines & Functions & Lines/func. & Del.\ share & New file \\
\midrule
\rowcolor{gray!12} change propagation & 170 & 3 & 114 & 7 & 15 & 0.30 & 16\% \\
new shared abstraction & 174 & 4 & 258 & 8 & 28 & 0.10 & 63\% \\
\rowcolor{gray!12} optimization clone & 124 & 4 & 92 & 8 & 10 & 0.40 & 16\% \\
parallel optimization & 78 & 4 & 152 & 7 & 20 & 0.20 & 27\% \\
\bottomrule
\end{tabularx}
\end{table}

This chain connects RQ3's characterization to RQ2's association.
The most common root cause is resolved by caching in 67\% of cases, a cache that serves several call sites is a new shared abstraction, and that pattern adds the most code and deletes the least, whereas RQ2 found the deleted-line share to be the one fix-level feature that separates the outcomes within agent and within repository (Section~\ref{sec:rq2:shape}).
Within the limits of those stratified comparisons, resolutions that remove work sit at the accepted end (allocation reduction, 69\%; the clone deletes 0.40 of its lines) and those that install a mechanism to postpone work sit at the rejected end (lazy evaluation, 40\%).

\section{RQ4: What evidence do agents provide to support their performance fixes?}
\label{sec:rq4}

\begin{tcolorbox}[enhanced, colback=gray!10, colframe=black!75, boxrule=0.8pt, arc=3pt, drop shadow={black!50!white}, left=5pt, right=5pt, top=3pt, bottom=3pt]
\textbf{Insight:} Agents attach tests to their performance fixes more often than the human fixes \citet{zhao2023perf} coded in 13 projects, but few of the tests check the performance property, and we detect no association between test kind and acceptance.
Among the merged fixes we re-executed, the tests that came with a fix could not have told a delivered improvement from a missing one.
A fix needs a performance test with a sized workload, and a reviewer needs to read what it asserts.
\end{tcolorbox}

Agents change a test in 37\% of their fixes, mostly by editing existing test files, and only 11\% of fixes carry a performance test or benchmark.
We detect no association between the kind of test and acceptance, and when we re-execute 30 merged fixes, 18 meet our delivery criterion, 9 show no significant gain or regress, and nearly half change behavior on inputs their tests never reach.
This section counts the test changes (Section~\ref{sec:rq4:often}), then reads what the changed tests check and relates the kind of test to acceptance (Section~\ref{sec:rq4:what}), and finally re-executes merged fixes (Section~\ref{sec:rq4:delivered}).

\subsection{Test Changes in Agent Fixes}
\label{sec:rq4:often}

\begin{table}[t]
\caption{Test changes in the 1,259 fixes with file data, by agent. A performance file is a test file whose name marks it as benchmark or performance code; file columns count touched test files by GitHub status (renamed, 1\%, omitted); case columns are the coders' reading of each fix with a test change (Table~\ref{tab:codes:rq4}; three fixes remove tests). Other pools Cursor, Claude Code and Jules.}
\label{tab:testchanges}
\footnotesize
\setlength{\tabcolsep}{3pt}
\begin{tabularx}{\textwidth}{>{\raggedright\arraybackslash}Xrrrrrrrrrr}
\toprule
 & & \multicolumn{2}{c}{Fixes with} & \multicolumn{4}{c}{Test files touched} & \multicolumn{3}{c}{Test cases} \\
\cmidrule(lr){3-4}\cmidrule(lr){5-8}\cmidrule(lr){9-11}
Agent & Fixes & test change & perf.\ file & n & mod. & added & rem. & added & mod. & both \\
\midrule
\rowcolor{gray!12} Copilot & 486 & 226 (47\%) & 44 & 786 & 57\% & 36\% & 5\% & 68\% & 14\% & 17\% \\
Codex & 438 & 135 (31\%) & 20 & 853 & 85\% & 15\% & 0\% & 37\% & 46\% & 17\% \\
\rowcolor{gray!12} Devin & 147 & 42 (29\%) & 4 & 133 & 44\% & 29\% & 27\% & 52\% & 31\% & 14\% \\
Other & 188 & 69 (37\%) & 14 & 451 & 69\% & 21\% & 10\% & 52\% & 19\% & 29\% \\
\midrule
\rowcolor{gray!12} All & 1,259 & 472 (37\%) & 82 & 2,223 & 69\% & 24\% & 6\% & 56\% & 25\% & 19\% \\
\bottomrule
\end{tabularx}
\end{table}
\begin{figure}[t]
\centering
\includegraphics[width=0.88\linewidth]{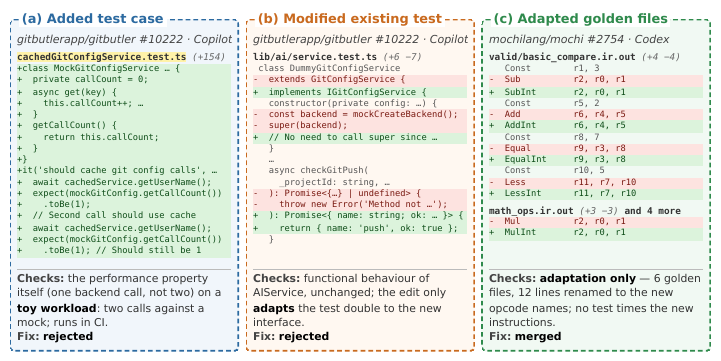}
\caption{Three kinds of test change, condensed from the diffs: (a) a test case added in a new file, (b) an existing test modified, and (c) golden files regenerated to the new output, which the coders record as an adaptation.}
\label{fig:testkinds}
\end{figure}
Agents change at least one test file in 37\% of the 1,259 fixes with file data, and 7\% touch a file whose name marks it as benchmark or performance code (Table~\ref{tab:testchanges}).
This count reads file names only, and Section~\ref{sec:rq4:what} reads what the tests check.
\citet{zhao2023perf} report a test change in 15\% of the 570 human-written performance fixes they coded in 13 projects; the agent fixes in our set change a test more often (37\%), though the two corpora differ in source, projects and coders (Section~\ref{sec:bg:human}).

The share differs by agent, from 47\% (Copilot) to 29\% (Devin), and the median test-changing fix touches one test file for Copilot and Devin, two for Codex.

Of the touched tests, most are edits to existing test files rather than new ones, since 69\% of the 2,223 test files that the 472 fixes touch are modified and 24\% are added.
The count is skewed, though: three fixes touch more than 100 test files each (222 in evmts/guillotine, 186 in department-of-veterans-affairs/vets-api and 107 in mochilang/mochi), and two of them hold 76 of the 123 removed files.
The coders instead read each fix at the level of test cases (Table~\ref{tab:codes:rq4}; Figure~\ref{fig:testkinds} shows the three kinds).
By that reading, 56\% of the 472 fixes add new test cases, 25\% modify existing ones, 19\% do both and 1\% (3) remove tests (rounding sums to 101\%).
The agents differ here as well: 85\% of the test files Codex touches are modified, most of them regenerated golden files in mochilang/mochi (Figure~\ref{fig:testkinds}c), whereas 36\% of the test files Copilot touches are new.

\subsection{The Impact of Tests on Acceptance}
\label{sec:rq4:what}
\label{sec:rq4:acceptance}

The changed tests check functional behavior in about half of the fixes that change them, only 11\% of all fixes include a performance test or benchmark, and we detect no association between the kind of test and acceptance.
In the 472 fixes with test changes, the tests assert functional behavior in 51\% and a performance quantity in 20\%; the remaining fixes counted under a performance test add a benchmark without an assertion.
In 30\% the test change includes an adaptation of existing tests, in 25\% the tests reproduce the inefficiency as a regression guard, and in 11\% the fix adds a benchmark without an assertion.
By agent, Codex's test changes include such adaptations in 52\% against 19\% for Copilot's.

\textit{Across the 1,259 fixes with file data, only 11\% include a performance test or benchmark}, while 16\% change functional tests only, 10\% only adapt tests and 63\% change no test, so the first three kinds together are the 37\% of Section~\ref{sec:rq4:often}.
Of the 140 fixes with a performance test or benchmark, 66 use a sized workload, 49 a toy one and 23 a parameterized one (2 use none), and 91 run in CI.

A performance test with a sized workload turns the claimed speed-up into a bound that CI checks on every run.
In calcom/cal.com (PR 21371), Devin replaced a scan over all slot boundaries in \texttt{getSlots} with a sorted lookup that stops at the first match, and the added test builds up to 2,000 date ranges, asserts ``expect(executionTimeInMs).toBeLessThan(2000)'' and runs in CI.
A reviewer suggested a one-second bound, and the fix was merged after 1.1 hours.
Because the PR reports a drop from about 6 seconds to 70 ms, the test would catch a return to the old loop but not a regression that loses most of the gain.

A toy workload with a call-count assertion checks that a cache is in place and leaves the slow path untested.
In gitbutlerapp/gitbutler (PR 10222), where each branch card made two calls to fetch the global git user name and email, Copilot added a \texttt{CachedGitConfigService} whose test calls \texttt{getUserName} twice on a mock and asserts ``expect(mockGitConfig.getCallCount()).toBe(1)''.
The test never renders a branch list, so it cannot show that a page with many branches issues fewer calls.
The maintainer closed the PR within 90 minutes because the repository's query system already caches (Section~\ref{sec:rq3:depth}), and no test in the PR could settle where the cache lives.

Codex's adaptations are mostly regenerated snapshots; 44 of its 70 are in mochilang/mochi.
In PR 2754 there, Codex made the virtual machine emit type-specialized integer and float instructions, and its only test change renames opcodes in six golden IR files (\texttt{Add} to \texttt{AddInt}), which shows the new instructions are emitted while no test times them.
Such a fix counts toward the 37\% with a test change yet adds no check.

\begin{figure}[t]
\centering
\includegraphics[width=\linewidth]{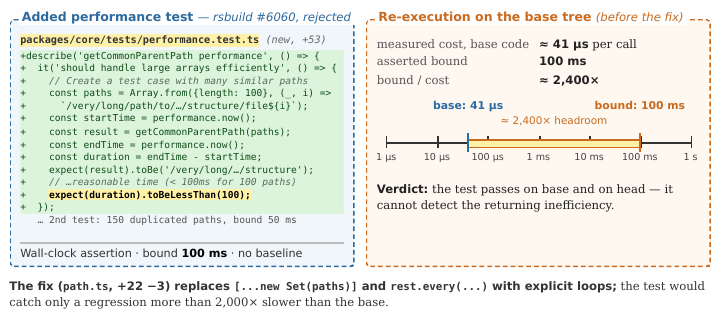}
\caption{A performance test whose bound sits far above the cost it guards (web-infra-dev/rsbuild PR 6060, condensed). The added test asserts 100 ms for 100 paths; the unfixed code takes about 41 \(\mu\)s per call, so the test passes on both trees.}
\label{fig:testbound}
\end{figure}

Fixes with test changes are accepted at 55\% and fixes without at 58\% (p = 0.49).
By kind, acceptance is 51\% with a performance test, 54\% with functional tests only, 60\% with adaptation only and 58\% with no test (p = 0.53).
Both comparisons hold within each agent and with or without third-party review.
For example, acceptance by kind is 49/54/62/62\% for fixes with review and 55/55/58/56\% for fixes without.
We also refit M0 on the 1,102 modeled fixes with the four test kinds in place of its tests indicator, keeping agent, size, attached measurement, third-party review and stars.
In that model every test-kind OR lies between 0.95 and 1.22 (p \textgreater{} 0.37), while size, third-party review, agent and stars keep their effects.
Several tests that the fixes added also pass on the unfixed base (Section~\ref{sec:rq1:validity}), so they would not detect a returning inefficiency.
For example, in web-infra-dev/rsbuild (PR 6060), Copilot added a \texttt{performance.test.ts} that requires \texttt{getCommonParentPath} to handle 100 paths within 100 ms, and in our re-execution the test also passed against the base code (Figure~\ref{fig:testbound}).
The base takes about 41 \(\mu\)s per call on 100 paths, so the bound sits more than 2,000 times above the cost it is meant to guard.

\subsection{Accepted Fixes That Do Not Deliver}
\label{sec:rq4:delivered}

\begin{figure}[t]
\centering
\includegraphics[width=0.84\linewidth]{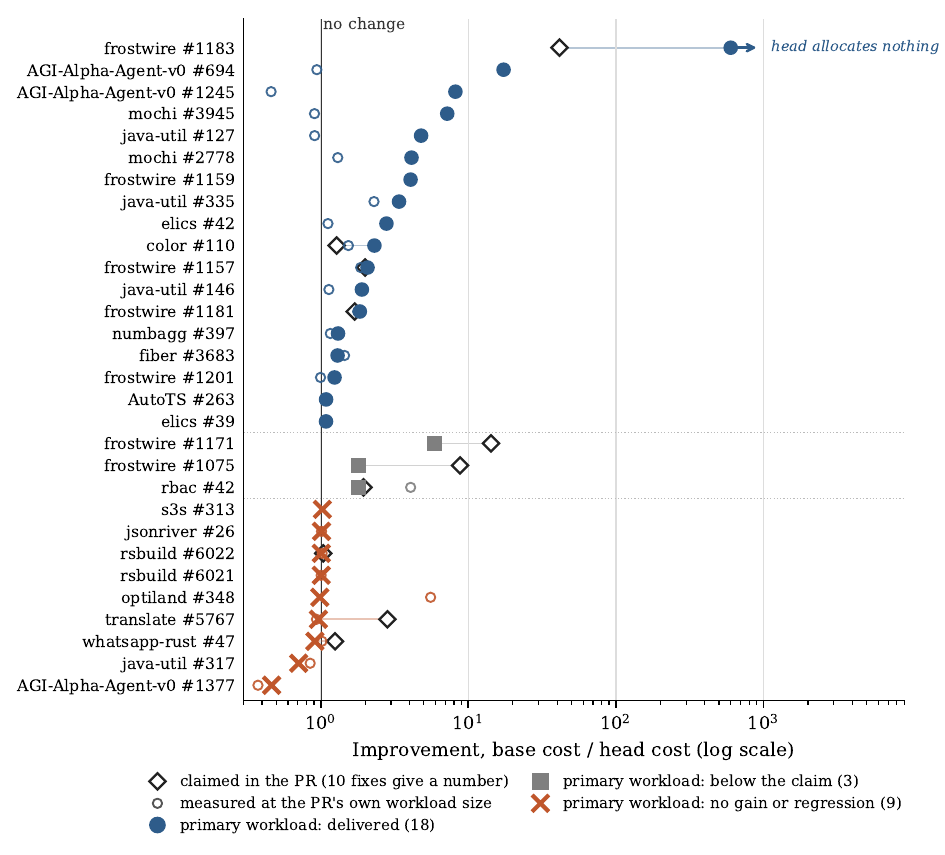}
\caption{Claimed and measured improvement of the 30 re-executed merged fixes: filled markers, the base-to-head cost ratio at the primary workload (which sets the verdict); small open circles, the ratio at the PR's own test size; open diamonds, the claimed improvement where the PR gives a number. frostwire 1183 removes an allocation entirely, so its ratio is unbounded.}
\label{fig:delivered}
\end{figure}

\textbf{Eighteen of the 30 merged fixes meet our delivery criterion at the primary workload, three improve but fall short of the claimed gain, nine show no significant gain or regress, and the tests they came with would not have told the difference.}
Of the nine that do not deliver, six show no gain of at least 5\% and three are slower (Figure~\ref{fig:delivered}).
The median ratio of base to head cost is 1.8\(\times\).
Ten fixes claim a number, and the measurement reaches it in four (frostwire 1157, 1181 and 1183, color 110), falls short of it in three (frostwire 1075 and 1171, rbac 42) and shows no gain or a loss in three (translate 5767 and rsbuild 6022 claim 2.8\(\times\) and 4\%, and whatsapp-rust 47 claims 1.25\(\times\) and allocates about 10\% more).
The gain also depends on the workload.
Eight fixes are slower than the base at one of the three sizes, and five of the eighteen that deliver gain only at the larger sizes.
In jdereg/java-util (PR 127), a binary search over a compact map's keys runs 9\% slower than the linear scan at the PR's own four keys and 4.8\(\times\) faster on 600.

The two arms are not comparable, so we report them side by side and draw no conclusion from their difference.
The rejected arm was selected for an attached measurement, ran three repetitions of a workload the executing agent mostly constructed, and used the reproduction criterion of Eq.~\ref{eq:repro}; the merged arm was selected for a test change and for feasibility, ran 12 forks at three sizes, and used the delivery criterion of Section~\ref{sec:design:rq4:metrics}, and only 10 of its 30 fixes state a number.
A difference between 18 of 30 and 6 of 23 therefore mixes the maintainers' decision with the sampling and the measurement standard, and it cannot show whether maintainers recognize fixes that deliver.
Table~\ref{tab:delivered} gives the counts by arm and by whether the fix's tests assert a performance property.
Within the merged arm, 3 of the 9 fixes whose tests carry a performance assertion meet the criterion and 15 of the 21 without one do (Fisher's exact p = 0.10); with nine and twenty-one fixes the comparison cannot rule out a difference in either direction, so we do not read it as evidence that a performance assertion predicts delivery, or that it does not.

\begin{table}[t]
\caption{Re-executed claims by arm. The rejected arm is the pilot of Section~\ref{sec:rq1:validity}; the merged arm is read at the primary workload (Section~\ref{sec:design:rq4:metrics}). ``Partly'' is a partly reproduced claim in the rejected arm and a gain below the claimed number in the merged arm.}
\label{tab:delivered}
\small
\setlength{\tabcolsep}{4pt}
\begin{tabular}{@{}llrrrr@{}}
\toprule
Arm & Tests in the fix & Run & Reproduced & Partly & Not \\
\midrule
\rowcolor{gray!12} Rejected & all & 23 & 6 & 4 & 13 \\
Merged & with a performance assertion & 9 & 3 & 2 & 4 \\
\rowcolor{gray!12} Merged & without & 21 & 15 & 1 & 5 \\
Merged & all & 30 & 18 & 3 & 9 \\
\bottomrule
\end{tabular}
\end{table}

The tests in the merged fixes rarely check the improvement, and several fail on the code that was merged.
The PR's own tests pass unchanged on the base tree in 17 of the 28 fixes in which they run there, so they would pass with or without the fix, and only 2 of the 30 fail on the base for a performance reason.
In four fixes (java-util 127, frostwire 1181 and 1183, AGI-Alpha-Agent-v0 1377) the PR's tests fail on the merged head itself.
The PR's tests execute every performance-relevant changed line in 10 of the 30 fixes, with a median diff coverage of 0.89, and the tests we generated reach full coverage in 28.
Coverage does not buy delivery: 8 of the 10 fully covered fixes fail the delivery criterion against 4 of the 20 partly covered ones (p = 0.004), because the fully covered fixes are mostly one- to five-line micro-optimizations whose gain disappears under the JIT or inside untouched work, so diff size confounds the comparison.

The generated tests also show that the head often behaves differently from the base on inputs the PR did not test.
Their outputs differ in 24 of the 30 fixes.
In ten the difference is the intended effect or has no consequence for callers, for example a rate limiter that evicts stale keys or a parser that now throws where the base hung.
In the other fourteen the merged fix introduces a functional change its description does not mention, and in ten of them the head fails, or returns a different result on inputs within the changed function's intended use, where the base did not.
The java-util binary search misses a key inserted in descending order, which the repository fixed later; the rbac permission check (PR 42) lets a child role's failed conditional permission shadow a parent's grant and lets glob queries match inherited operations; the mochi compiler (PR 3945) emits a string where a hashed left join should emit a number; the numbagg covariance (PR 397) switches to a one-pass formula whose relative error grows to 0.13 at an offset of \(10^6\); the AGI-Alpha-Agent-v0 rate limiter (PR 1245) lets a throttled client through once more than 1,024 clients are active; and the elics entity store (PR 42) throws on the 32nd component type.
None of these inputs is in the PR's tests, and where the reports record it, the PR's own benchmark ran on data that never reaches the differing input.

Read against Section~\ref{sec:rq4:what}, the tests in the merged fixes do not carry the information a merge would need: we detected no association between a test in the fix and a merge, the two small groups with and without a performance assertion do not separate delivery, and the tests that were merged pass on the unfixed code in most fixes.
Nine of the 30 merged fixes show no significant gain or regress at the primary workload, and nearly half change behavior on inputs their tests never reach, so a merge records that a maintainer accepted the fix and not that the fix does what its description says.

\section{Discussion}
\label{sec:discussion}

\subsection{Implications}
\label{sec:discussion:implications}

\textit{For agent builders, the findings point to the shape of a fix and to the repositories that already merge the agent's work.}
The deleted-line share is the one fix-level signal that holds up within agent and within repository, and the pre-opening merge rate and the track record improve model fit more than any content block (Section~\ref{sec:rq2}).
Smaller fixes that remove code are associated with higher acceptance than large added mechanisms, although whether changing a fix's shape would change its outcome is untested.
The advice matters most for architectural-level fixes, where a new shared cache or helper is as common as a propagated interface change (Section~\ref{sec:rq3:scope}).

\textit{For maintainers, a measurement they can re-run would put fix-specific evidence into the merge decision.}
In our data, 61\% of rejections state no reason and reviewers ask for a measurement in 3\% (Section~\ref{sec:rq1:grounds}).
In the pilot, the claim reproduced in a quarter of the re-executed rejected fixes and failed in more than half, though most failures surfaced on workloads the executing agent constructed (Sections~\ref{sec:rq1:validity}, \ref{sec:threats}).
Among merged fixes, measured under a different protocol, 18 of 30 met our delivery criterion at the primary workload and 9 showed no significant gain or regressed, and the tests that came with them pass on the unfixed code in 17 of 28, so the merged artifact does not show that the claim was checked (Section~\ref{sec:rq4:delivered}).
A performance test in the fix plays the same role once it fails on the base and passes on the head.

\textit{Researchers who evaluate agents can measure the speed-up and the reach of a fix in addition to what its PR states.}
The re-execution protocol of Section~\ref{sec:pilot} costs one language container and at most thirty minutes per fix and resolved the large claims in 19 of the 23 fixes we ran (6 reproduced, 13 failed).
The merged arm's more intensive protocol surfaces failure modes the pilot cannot: with 12 interleaved forks at three workload sizes, 8 of 30 fixes are slower than the base at one size, and generated tests over the changed lines found an unmentioned functional change in 14 of 30, so an evaluation that reports one number on one workload can call a fix delivered when it regresses at the PR's own scale or changes behavior its description does not mention.
Reach needs the same care: an architectural-level agent fix changes one or two functions per file (Section~\ref{sec:rq3:depth}), which suggests one small edit pattern repeated across a codebase, so an evaluation of architectural-level capability should record per-file depth as well as file count.

\textit{For researchers who build on AIDev, a changed test is weak evidence that the claimed benefit was verified.}
Agents change tests in more than a third of fixes, yet only one fix in nine includes a test of the performance property (Section~\ref{sec:rq4:what}).
About a third of the performance tests and benchmarks use a toy workload, and some added tests pass on the unfixed base (Section~\ref{sec:rq4:acceptance}).
Studies that use AIDev should therefore code what a test checks before using test changes to measure verification.

\subsection{The Approval Funnel against Human PRs in the Same Repositories}
\label{sec:funnel}

\begin{figure}[t]
\centering
\includegraphics[width=0.86\linewidth]{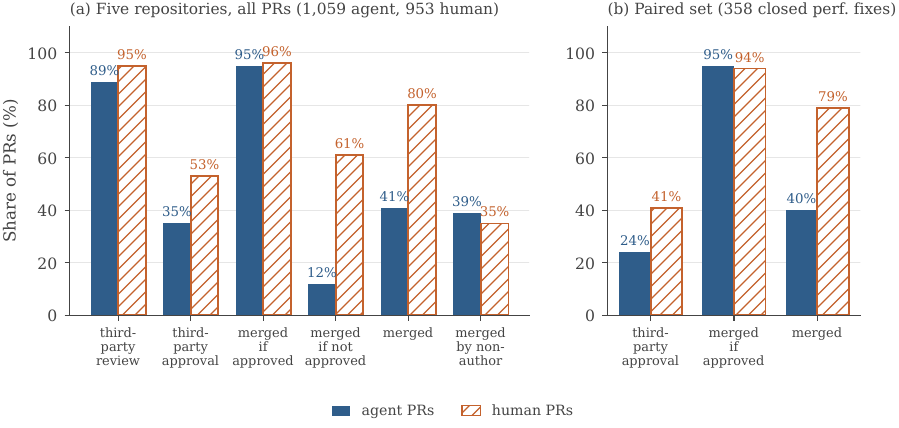}
\caption{The approval funnel of agent and human PRs: (a) all PRs opened January--June 2025 in the five repositories, by step (third-party review or comment, third-party approval, merge rate among approved and among unapproved PRs, overall merge rate, merges by someone other than the author); (b) the same steps in the paired set's 358 closed performance fixes.}
\label{fig:funnel}
\end{figure}
In RQ1 and RQ2, acceptance varies with the account model and the repository more than with the fix's coded properties.
To test whether this pattern is specific to agents, we compare agent and human PRs reviewed by the same maintainers in two sources.
The first source is the earlier paired set on AIDev v3, built with the pipeline of Section~\ref{sec:population} on both sides: 247 agent and 128 human performance fixes in the 810 repositories holding PRs from both, with AI-assisted PRs removed from the human side.
The second is a re-fetch of every PR opened January--June 2025 in the five repositories with the most PRs from both sides (airbyte, cal.com, crewAI, liam, onlook); removing bots and AI-written PRs \citep{duma2026reviews} from the 2,306 fetched leaves 1,059 agent and 953 human PRs.

\begin{table}[t]
\caption{The approval funnel in the five repositories (1,059 agent and 953 human PRs opened January--June 2025) and in the paired set (358 closed performance fixes). p is Fisher's exact test; the odds ratios hold the author's association, the size of the change and the repository fixed.}
\label{tab:funnel}
\small
\setlength{\tabcolsep}{4pt}
\begin{tabularx}{\textwidth}{>{\raggedright\arraybackslash}Xrrrrrr}
\toprule
 & \multicolumn{4}{c}{Five repositories} & \multicolumn{2}{c}{Paired set} \\
\cmidrule(lr){2-5}\cmidrule(lr){6-7}
Step & Agent & Human & p & OR & Agent & Human \\
\midrule
\rowcolor{gray!12} third-party review or comment & 89\% & 95\% & \textless{} 0.001 & & 52\% & 59\% \\
third-party approval & 35\% & 53\% & \textless{} 0.001 & 0.23 & 24\% & 41\% \\
\rowcolor{gray!12} merged, among approved & 95\% & 96\% & 0.39 & 0.50 & 95\% & 94\% \\
merged, among not approved & 12\% & 61\% & \textless{} 0.001 & & & \\
\rowcolor{gray!12} merged & 41\% & 80\% & \textless{} 0.001 & 0.17 & 40\% & 79\% \\
merged by someone other than the author & 39\% & 35\% & 0.052 & 0.87 & & \\
\bottomrule
\end{tabularx}
\end{table}

\textit{Agent PRs obtain third-party approval less often than human PRs, and once approved, the two merge at the same rate (Figure~\ref{fig:funnel}, Table~\ref{tab:funnel}).}
In the five repositories, agent PRs (almost all Devin) are merged 41\% of the time and human PRs 80\%, and the gap appears in every repository (16--58 points; pooled p \textless{} 0.001).
With author association, size and repository held fixed, agent PRs have an OR of 0.17.
Third-party review is similarly common, since 89\% of agent and 95\% of human PRs receive a review or comment from an account other than the author's.
This measure counts comments posted from a maintainer's account even when a tool wrote them: in crewAIInc/crewAI, 220 of the 283 agent PRs carry the maintainer account's ``This review was made by a crew of AI Agents'' comment, and in 162 it is the only third-party comment.
However, a third party approves 35\% of agent versus 53\% of human PRs.
After approval the gap closes, as 95\% of agent and 96\% of human PRs are merged (p = 0.39).
The same funnel appears in the 358 of the paired set's 375 fixes that are closed: merged 40\% versus 79\%, approved 24\% versus 41\%, merged after approval 95\% versus 94\%.

Human authors who merge their own unapproved PRs account for the rest of the gap.
Among human PRs with no third-party approval, 61\% are merged anyway, and the author merges 88\% of these, whereas 12\% of unapproved agent PRs are merged.
Counting only merges by someone other than the author, the two populations are level (39\% versus 35\%, Fisher p = 0.052; adjusted OR 0.87, p = 0.25).

In airbytehq/airbyte the gap comes almost entirely from the author's merge button: a third party approves only 17 of its 162 human PRs, yet 152 are merged, the author merging 136 of the 145 unapproved; of its 169 agent PRs, 72 are approved and 63 of those merged, against 8 of the 97 unapproved.

In crewAIInc/crewAI the gap arises at approval itself, since a third party approves 31 of the 283 agent PRs and 205 of the 284 human PRs.
For example, Devin's PR 2799 deletes a two-line append in \path{agent_utils.py} that, per its description, duplicated tool results in the LLM prompt.
Two minutes after the PR opened, the collaborator account joaomdmoura posted a comment beginning ``This review was made by a crew of AI Agents''; no approval followed, and eight days later Devin closed the PR for inactivity.
Four weeks later, the developer lucasgomide opened PR 2964, which deletes the same two lines; another contributor approved it seven minutes later, and the author merged it 14 minutes after opening.
The other two agent performance fixes in crewAI (PRs 2281 and 3077) followed the same course: the same deletion merged under a human author and stalled at approval under a bot account.

The approval step also matches the account-model split of RQ1 (Section~\ref{sec:rq1:outcome}).
Where an operator holds the merge button (Codex, Cursor, Claude Code), agent fixes are accepted like a maintainer's own work.
However, where a bot account holds the PR (Copilot, Devin, Jules), acceptance rises with the repository's track record with that agent (Section~\ref{sec:rq2:context}).
These comparisons locate the lower acceptance of agent PRs at the approval step; what an approver weighs there is a question for interview work.

\section{Threats to Validity}
\label{sec:threats}

\textbf{Construct validity.}
Labels and codes come from codebook coding, by language models for the candidates that passed the text filter and then by the authors for every field of the model-positive set.
In each pass two coders coded independently and a third resolved disagreements; the two models agree on the verdict with a Cohen's \(\kappa\) of 0.86, the two authors with 0.91.
The agreement of language models with human annotators has been measured on text-annotation tasks \citep{gilardi2023chatgpt}, and the model stage only selects what the coders read.
The blind-pair judge and the re-execution agent are Claude Opus models, and 60 of the 1,262 fixes come from Claude Code, so a same-vendor bias in judging that agent's fixes cannot be ruled out for those two instruments.
Two further threats apply to the judge: the head commit of a merged fix can include changes made in response to review, so the judge may read a cleaner fix than the maintainer first saw, and these public PRs may lie in its training data, so a memorized outcome cannot be ruled out.
The authors' coding is exposed the same way, since the PR page shows the outcome the coders were not blind to.
The HLP text filter was derived from human-written issue reports; an audit of 200 PRs under its threshold found three performance fixes, which together with the screen's cross-validated recall puts its miss rate at roughly a tenth of the population, and every population share is conditioned on the filter.
The model coding stage between the filter and the authors was audited on a 1-in-10 sample of its 6,047 rejections (Section~\ref{sec:population}), which bounds its false negatives without eliminating them.
Function counts rely on git's hunk-header heuristic and are therefore reported as a lower bound.
The production-file classifier also counts a few non-code files (a .gitignore, a minified source map) as production files, which inflates the raw multi-file count that the coded scope corrects.
The hype-word list overlaps performance vocabulary and so favors a difference; none appeared within agents.

\textbf{Internal validity.}
The pre-opening merge rate uses only decisions made before the fix opened, but it reflects the same maintainers' or operators' habits, so it measures how the repository has treated agent PRs rather than an independent cause; for operator accounts, where the operator merges their own agent's work, it partly measures that operator's own habit.
The snapshot variant also counts later decisions and fits better for exactly that reason, which is why it stays a robustness check.
The track record, as a raw count, mixes prior success with exposure and repository activity, so it measures history rather than a trust mechanism.
For agent bot accounts, an author association of NONE coincides with an empty track record (Section~\ref{sec:design:rq1:approach}), so the two variables are partly collinear in M4.
Operator accounts (Codex, Cursor, Claude Code) and bot accounts (Copilot, Devin, Jules) are therefore reported separately.
Devin's inactivity template makes the \texttt{closed\_by} field unreliable, so we rely on the coded closing reason.

\textbf{Conclusion validity.}
The study runs many significance tests, so a single difference may arise by chance.
The central results rest on the same pattern appearing under several methods: the deleted-line share differs pooled (Mann--Whitney), within each agent, and across the 74 repositories that hold both outcomes (Wilcoxon), and the repository variables hold in the pre-opening bands, in the regressions, and in the snapshot robustness check.
However, these analyses are dependent robustness checks on one dataset rather than independent replications, and the subgroup and category comparisons are exploratory rather than confirmatory.
Dropping the two most concentrated repositories (mochilang/mochi with 91 consensus fixes, Smart-Cleaner-for-Android with 45) moves pooled acceptance to 55\%.
In contrast, the content differences that do appear are isolated, namely lazy or deferred evaluation and the file-count gradient (each at p = 0.004) and architectural-level scope (p = 0.044).
They would weaken under any correction, so we read them as suggestive.
The deleted-line share enters the acceptance models as one block among four, so its coefficient is conditional on the blocks entered before it.
Fixes in one repository share maintainers, so the acceptance models cluster standard errors by repository.

\textbf{External validity.}
AIDev v4 covers six agents in repositories over 100 stars up to October 2025, and the 157 open fixes may since have closed.
The re-execution pilot has 23 fixes, and for 16 of them the executing agent constructed a workload on the changed path rather than running one the PR named.
It shows that non-reproduction is common among rejected fixes with an attached measurement, and how it arises, but gives no population rate.
The merged arm of Section~\ref{sec:rq4:delivered} was selected by judgment for fixes that a deterministic in-process harness can drive and that build with the toolchains we had, which favors small, self-contained fixes, whose gains most easily vanish under a JIT or inside untouched work, so its non-delivery rate (9 of 30) may overstate the population's.
The agent PRs in the case repositories of Section~\ref{sec:funnel} are almost all Devin's, but the funnel replicates on the paired set.
The review-or-comment rate in that section counts AI-written reviews posted from a maintainer's account as third-party review, so it overstates human attention in at least one repository, and the silent-decision classification is conservative for the same reason; the approval and merge steps do not depend on comment text.

\textbf{Reliability.}
A re-run of the model labeling stage, the pair judge or the executing agent may not reproduce the same output, so we release the data, codebooks, coding sheets, model outputs, container results and scripts.

\section{Conclusion and Future Work}
\label{sec:conclusion}

Whether an agent's performance fix is merged varies with the agent's track record in the repository, the repository's pre-opening merge rate on its other agent PRs, and whether the fix removes or adds code.
We detect no association between acceptance and the fix's coded content, description, tests or measurements.
Evidence about the fix itself is rarely requested, and when we re-executed fixes, the rejected claims mostly failed and 9 of 30 merged fixes showed no significant gain or regressed.

Maintainers who re-run a fix's claimed speed-up before deciding would put evidence about the fix itself into the decision; because the merge outcome is confounded with the repository's prior treatment of the agent, studies that use acceptance as a quality measure need a within-repository control.

Four limits set the next steps.
First, the findings are associations in one snapshot of AIDev v4; whether changing a fix's shape would change its outcome is untested, and a controlled study that, with the maintainers' consent, submits the same optimization in a removing and an adding form to the same repositories would settle it.
Second, the two re-execution arms are small, measured under different protocols, and the merged arm was selected by feasibility; running both under the three-size protocol of Section~\ref{sec:rq4:delivered} on a random sample would give comparable delivery rates and show whether an agent-chosen workload understates them.
Third, the funnel comparison rests on five repositories and one paired set; extending it to repositories where humans and several agents submit fixes side by side would show whether the approval step treats every bot account alike.
Fourth, the 157 open fixes and agents released after October 2025 sit outside the snapshot; a refresh would show whether a repository's treatment of agent PRs moves as its track record grows.

\section*{Declarations}

\bmhead{Data and code availability}
Data, codebooks, coding sheets, model outputs, container results and scripts are available for review at [upon acceptance]; a DOI archive follows on acceptance.

\bmhead{Use of generative AI}
The authors used language models as study instruments (Section~\ref{sec:design}) and LLM-based tools in drafting and editing; they reviewed and verified all content and take full responsibility for it.

\bmhead{Conflict of interest}
The authors declare no conflict of interest.

\bibliography{references}

\end{document}